\documentclass[aip,pop,reprint]{revtex4-1}

\usepackage{graphicx}
\usepackage{dcolumn}
\usepackage{bm}
\usepackage[utf8]{inputenc}
\usepackage[T1]{fontenc}
\usepackage{mathptmx}
\usepackage{etoolbox}
\usepackage{cancel}
\usepackage{hyperref}
\usepackage{subfigure}
\usepackage{tabularx}
\usepackage{amsmath}
\usepackage{amsfonts}
\usepackage{amssymb}

\makeatletter
\def\@email#1#2{%
 \endgroup
 \patchcmd{\titleblock@produce}
  {\frontmatter@RRAPformat}
  {\frontmatter@RRAPformat{\produce@RRAP{*#1\href{mailto:#2}{#2}}}\frontmatter@RRAPformat}
  {}{}
}%
\makeatother

\begin{document}


\title[A Subgrid Model for Electron-Scale Turbulent Transport in Global Ion-Scale Gyrokinetic Simulations of Tokamak Plasmas]
{A Subgrid Model for Electron-Scale Turbulent Transport in Global Ion-Scale Gyrokinetic Simulations of Tokamak Plasmas}

\author{S. Tirkas}
\affiliation{Department of Physics, University of Colorado at Boulder, Boulder, CO 80309, USA}
\email{stefan.tirkas@colorado.edu.}
\author{Y. Chen}%
\affiliation{Center for Integrated Plasma Studies, Boulder, CO 80309, USA}
\author{S. Parker}
\affiliation{Department of Physics, University of Colorado at Boulder, Boulder, CO 80309, USA}
\affiliation{Renewable and Sustainable Energy Institute, Boulder, CO 80303, USA}

\date{\today}

\begin{abstract}
A subgrid electron-temperature-gradient (ETG) model is demonstrated here which averages local electron-scale turbulence over intermediate scales in space and time to
include in global ion-temperature-gradient (ITG) simulations. This approach results in ion-scale equations which incorporate the electron heat transport from ETG
turbulence and effects of electron-scale turbulence on the ion scale. Flux-tube ETG Cyclone Base Case simulations are performed at different radial locations and a
kinetic form of the flux is added to global ion-scale simulations as a source term. Analytic radial profiles of ETG heat flux are constructed and compared to flux-tube
results at multiple radial locations. Different ratios of ITG to ETG heat flux levels are considered and the results of capturing ETG heat transport in global
ion-scale simulations are discussed. Potential coupling of the ETG streamer potential and intermediate-scale zonal flows to the ion scale is further addressed.
\end{abstract}


\maketitle

\section{Introduction\label{sec:intro}}

It is now well established that turbulence is responsible for the anomalous transport of heat and particles in tokamaks. This turbulence exists in the form of
microinstabilites at ion and electron gyroradius scales which are driven by strong gradients in the equilibrium plasma profiles. Gyrokinetic simulations of ion
gyroradius scales are currently able to confidently predict ion transport and power spectra in experiment; however, they can often underestimate electron thermal
transport.\cite{LmodeETG_Howard13,MultiChannel_White2013} The electron-temperature-gradient (ETG) mode is a primary candidate to account for this excess heat loss
and is characterized by radially extended `streamers' at electron gyroradius scales.\cite{Gene1_Jenko2000,ETG_Dorland2000,ETG_Jenko2002}

The role of ETG turbulence has been studied in various tokamak scenarios\cite{ExptETG_Gurchenko2010,OverviewMAST_Chapman2015,ETG_NSTX_Ren2017,AsdexETG_Ryter2019,
ECH_PFPO1_Kiefer2021} and is particularly important in cases of marginal ion-scale turbulence.\cite{ReviewETG_Ren2024} Specifically for ITER, alpha-particle and
electron cyclotron heating effects are expected to drive meaningful ETG turbulence levels, leading to important multiscale dynamics.\cite{MultiscaleIBS_Holland2017,
Multiscale3_Howard2018} To better understand interactions between the disparate scales, local multiscale simulations of core ITG and ETG turbulence have garnered
much interest.\cite{MultiJenko_2004,MultiscaleCBC_Maeyama2015,Multiscale2_Howard16,MultiscaleIBS_Holland2017,Multiscale3_Howard2018,MultiscaleJet_Bonanomi2018,
MultiscaleIBS_Howard2021} In general, turbulent spectra are distinctly scale-separated, and cross-scale interactions lead to changes in steady-state transport
levels which can better predict experimental losses for both species. A recent overview of multiscale simulation results can be found in Ref.
\onlinecite{MultiscaleOverview_Maeyama2024}.

As multiscale simulations require resolving electron gyroradius scales, the sizes of simulation domains become limited. Consequently, reduced modeling of electron-scale
turbulence is particularly valuable for whole-device modeling efforts in future burning plasma experiments. Previous work has considered the importance of cross-scale
interactions,\cite{OldMulti_Itoh02,MultiJenko_2004,OldMulti_Holland05} while more recent efforts have developed reduced models for pedestal ETG transport
\cite{ReducedETG_Hatch22} and multiscale quasilinear saturation rules.\cite{MultiQL_Staebler16} Additionally, a scale-separated model of coupled gyrokinetic
equations\cite{MultiscaleModel_Hardman2019} has shown that ion-scale turbulence influences electron-scale dynamics through parallel-to-the-field shearing which suppresses
the ETG growth rate.\cite{MultiscaleModel2_Hardman2020}

The goal of this work is to account for heat losses due to ETG turbulence in global ion-scale simulations and to probe the effects of ETG turbulence on the ITG background.
The paper is outlined as follows. Section \ref{sec:model} describes a theoretical model which accounts for electron-scale effects in global ion-scale gyrokinetic
simulation. Local ITG and ETG simulations are carried out in GENE to test for a valid scale-separated scenario and the results are described in section \ref{sec:gene}.
Section \ref{sec:gem} then describes the global ITG simulation in GEM. This is followed by the results of including a kinetic source term that accounts for excess
electron-scale thermal losses due to ETG turbulence in GEM. Section \ref{sec:disc} concludes with future plans to couple the ETG streamer potential and intermediate-scale
zonal flows\cite{ETG-ZF_Colyer2017,MultiscaleIBS_Holland2017,ETG-ZF_Tirkas2023} found in ETG simulations to the ion scale.

\section{Subgrid ETG Model\label{sec:model}}

The gyrokinetic framework for modeling microturbulence in tokamak plasmas assumes an expansion in the parameter $\epsilon = \rho/a \ll 1$, where $\rho$ is the species'
gyroradius involved in the generation of instabilities and $a$ is the device minor radius.\cite{gkDeriv_FriemanChen82,gkDeriv_Lee83,gkDeriv_Dubin83} This in part allows
for separating dynamical equations between small-scale fluctuating quantities and background equilibrium quantities. A further subsidiary expansion can be made using the
small electron-to-ion mass ratio, $\sqrt{m_e/m_i} \ll 1$, to separate the dynamics of ion-scale (IS) and electron-scale (ES) instabilities, such as ITG and ETG modes.
\cite{MultiscaleModel_Abel13,MultiscaleModel_Hardman2019} Distinct equations can then be used to investigate the effects of coupling between the two scales. Note that
a lowercase `es' is instead used to refer to electrostatic simulations further on.

The primary assumption of the subgrid model is to take the electron-scale gyrokinetic equation as stand-alone, i.e. unaffected by ion-scale turbulence. Then local
ES turbulence is averaged over intermediate scales in time and space perpendicular to the field to include effects of ETG turbulence in global ITG simulations.
The amplitude of the ES flux-tube steady-state is varied in accordance with values reported in multiscale simulations involving core ITG and ETG turbulence and the
effect of new terms are either discussed further or investigated.

The subgrid model might further be incorporated with the multiscale model of Ref. \onlinecite{MultiscaleModel_Hardman2019} to perform self-consistent coupled simulations
of ITG and ETG turbulence. As the theory in Ref. \onlinecite{MultiscaleModel_Hardman2019} makes various assumptions about the electron-scale turbulence, the final equations
focus on the effects of ITG turbulence on the electron scale. The goal here is to recreate effects of electron-scale turbulence in multiscale simulations by using minimal
assumptions and incorporating ES effects directly from flux-tube simulation. The intermediate-scale averaging procedure offers a reduced approach for including ES
turbulent effects, making it practical for use in global ion-scale simulations. The accuracy of the subgrid model in capturing cross-scale interactions can be validated
by comparing to local IS simulations that directly include ES turbulence.

In focusing on the combined effects of ITG and ETG turbulence, the subgrid model includes only electrostatic effects. The governing gyrokinetic Vlasov equation takes the
form
\begin{equation}\label{gkVlasov}
\begin{aligned}
   \frac{\partial\delta f}{\partial t} &+ \left(v_{\parallel}\textbf{b} + \textbf{v}_D\right)\cdot\nabla\delta f
   + \frac{1}{B}\langle\delta\textbf{E}\rangle_{\alpha}\times\textbf{b}\cdot\nabla\delta f \\
   &= -\frac{1}{B}\langle\delta\textbf{E}\rangle_{\alpha}\times\textbf{b}\cdot\nabla f_0 + q(v_{\parallel}\textbf{b}
    + \textbf{v}_D)\cdot\langle\delta\textbf{E}\rangle_{\alpha}\frac{f_0}{T},
\end{aligned}
\end{equation}
where $\langle\dots\rangle_{\alpha}$ represents a gyro-phase angle average. For simplicity, the gyroaverage notation is dropped from now on. Collisions are ignored
assuming low collision frequency in the core.

Fluctuations of the perturbed distribution function and electrostatic potential are split into IS and ES terms
\begin{equation}\label{spectra}
\begin{aligned}
   \delta f &= \delta f' + \delta f_{\textrm{ES}}, \\
   \delta\textbf{E} &= \delta\textbf{E}_{\textrm{IS}} + \delta\textbf{E}_{\textrm{ES}},
\end{aligned}
\end{equation}
where $\delta f'$ retains new components that may be generated in the original IS distribution function due to the inclusion of ES $\textbf{E}$x$\textbf{B}$ effects. In
general this could also include cross-scale energy cascading effects, though this is ignored with the stand-alone ES assumption. An averaging procedure over intermediate
mesoscales in time and in space perpendicular to the field, $\tau_{\textrm{m}}$ and $l_{\textrm{m}}$, can then be introduced, where the intermediate values lie between IS
and ES scales defined by $a \gg \rho_i \gg \rho_e$ and $\tau \gg a/v_{th,i} \gg a/v_{th,e}$. Here, $\tau$ is the transport timescale and $v_{th}$ is the particle thermal
velocity. The specific choice of intermediate-scale values is described further in section \ref{sec:gene} in reference to the ES flux-tube simulation results.

Mesoscale averaging the components of \eqref{spectra} results in retaining only IS fluctuations:
\begin{equation}\label{avgSpectra}
\begin{aligned}
   \langle\delta f'\rangle_{\textrm{m}} = \delta f_{\textrm{IS}},&\;\;\; \langle\delta f_{\textrm{ES}}\rangle_{\textrm{m}} = 0, \\
   \langle\delta\textbf{E}_{\textrm{IS}}\rangle_{\textrm{m}} = \delta\textbf{E}_{\textrm{IS}},&\;\;\; \langle\delta\textbf{E}_{\textrm{ES}}\rangle_{\textrm{m}} = 0.
\end{aligned}
\end{equation}
Here, $\langle\dots\rangle_{\textrm{m}}$ represents a mesoscale average in space perpendicular to the field and time. It is assumed the mesoscale average of $\delta f'$
recovers the IS distribution function. As the electron scale is stand-alone, the ES dynamics are described by the gyrokinetic equation
\begin{equation}\label{gkETG}
\begin{aligned}
   \frac{\partial\delta f_{\textrm{ES}}}{\partial t} &+ \left(v_{\parallel}\textbf{b} + \textbf{v}_D + \frac{1}{B}\delta\textbf{E}_{\textrm{ES}}\times\textbf{b}\right)
   \cdot\nabla\delta f_{\textrm{ES}} \\
   &= -\frac{1}{B}\delta\textbf{E}_{\textrm{ES}}\times\textbf{b}\cdot\nabla f_0 + q(v_{\parallel}\textbf{b} + \textbf{v}_D)\cdot\delta\textbf{E}_{\textrm{ES}}\frac{f_0}{T}
   + S_{\textrm{ES}}.
\end{aligned}
\end{equation}
A source term, $S_{\textrm{ES}}$, has been explicitly included to ensure a steady-state consistent with the flux-tube approximation. To obtain an equation for the remaining
scales, the total fluctuating quantities are substituted into \eqref{gkVlasov} and then the ES equation subtracted, resulting in
\begin{equation}\label{gkTotal}
\begin{aligned}
   \frac{\partial\delta f'}{\partial t} &+ \left(v_{\parallel}\textbf{b} + \textbf{v}_D + \frac{1}{B}\delta\textbf{E}_{\textrm{IS}}\times\textbf{b}\right)\cdot\nabla\delta f'
   \\ &+ \frac{1}{B}\delta\textbf{E}_{\textrm{ES}}\times\textbf{b}\cdot\nabla\delta f' + \frac{1}{B}\delta\textbf{E}_{\textrm{IS}}\times\textbf{b}\cdot\nabla\delta f_{\textrm{ES}} + 
   S_{\textrm{ES}} \\
   &= -\frac{1}{B}\delta\textbf{E}_{\textrm{IS}}\times\textbf{b}\cdot\nabla f_0 + q(v_{\parallel}\textbf{b} + \textbf{v}_D)\cdot\delta\textbf{E}_{\textrm{IS}}\frac{f_0}{T}.
\end{aligned}
\end{equation}
Equation \eqref{gkTotal} is then averaged over the intermediate mesoscales to find the new IS gyrokinetic equation
\begin{equation}\label{gkITG}
\begin{aligned}
   \frac{\partial\delta f_{\textrm{IS}}}{\partial t} &+ \left(v_{\parallel}\textbf{b} + \textbf{v}_D + \frac{1}{B}\delta\textbf{E}_{\textrm{IS}}\times\textbf{b}\right)
   \cdot\nabla\delta f_{\textrm{IS}} \\
   &+ \langle\frac{1}{B}\delta\textbf{E}_{\textrm{ES}}\times\textbf{b}\cdot\nabla\delta f'\rangle_{\textrm{m}} + \langle\frac{1}{B}\delta\textbf{E}_{\textrm{IS}}\times\textbf{b}
   \cdot\nabla\delta f_{\textrm{ES}}\rangle_{\textrm{m}}
   + \langle S_{\textrm{ES}}\rangle_{\textrm{m}} \\
   &= -\frac{1}{B}\delta\textbf{E}_{\textrm{IS}}\times\textbf{b}\cdot\nabla f_0 + q(v_{\parallel}\textbf{b} + \textbf{v}_D)\cdot\delta\textbf{E}_{\textrm{IS}}\frac{f_0}{T}.
\end{aligned}
\end{equation}
The three new terms are grouped together on the middle line of \eqref{gkITG} for clarity. These new terms represent the averaged effects of electron-scale
turbulence in ion-scale simulations. The first term represents the additional guiding-center motion due to the ETG field. The second term is due to effects of the ITG
field on the ETG distribution function, which averages to zero according to \eqref{avgSpectra}. The final term, $S_{\textrm{ES}}$, is used to account for electron thermal
transport by ES turbulence in IS simulations.

\begin{figure*}
   \centering
   \subfigure[]{
   \label{fig:EqProfiles}
   \includegraphics[width=2.2in]{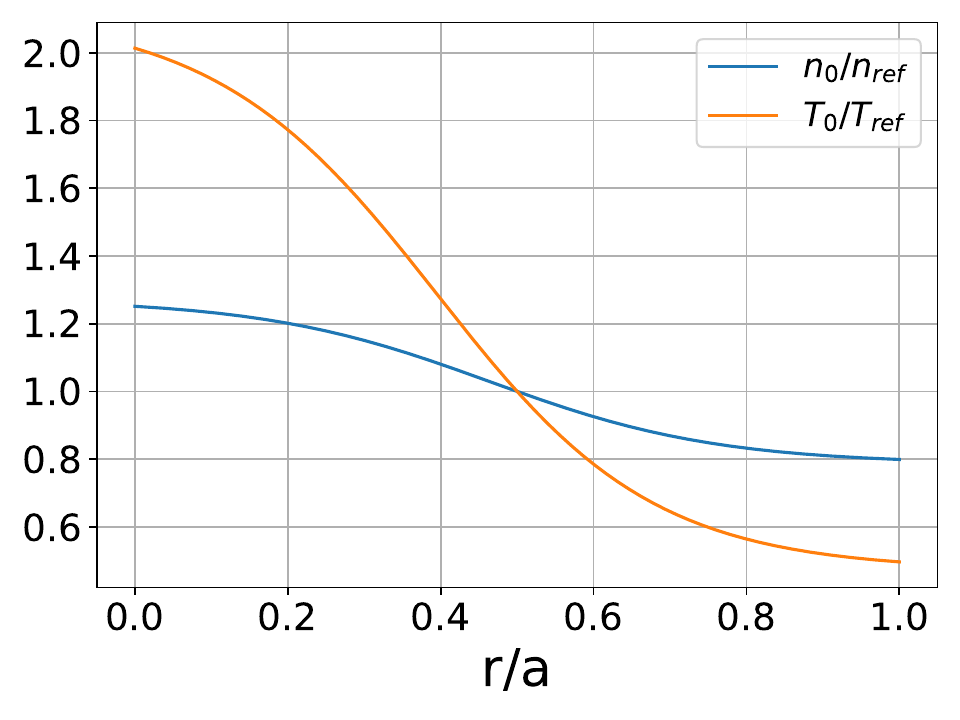}}
   \subfigure[]{
   \label{fig:GradProfiles}
   \includegraphics[width=2.2in]{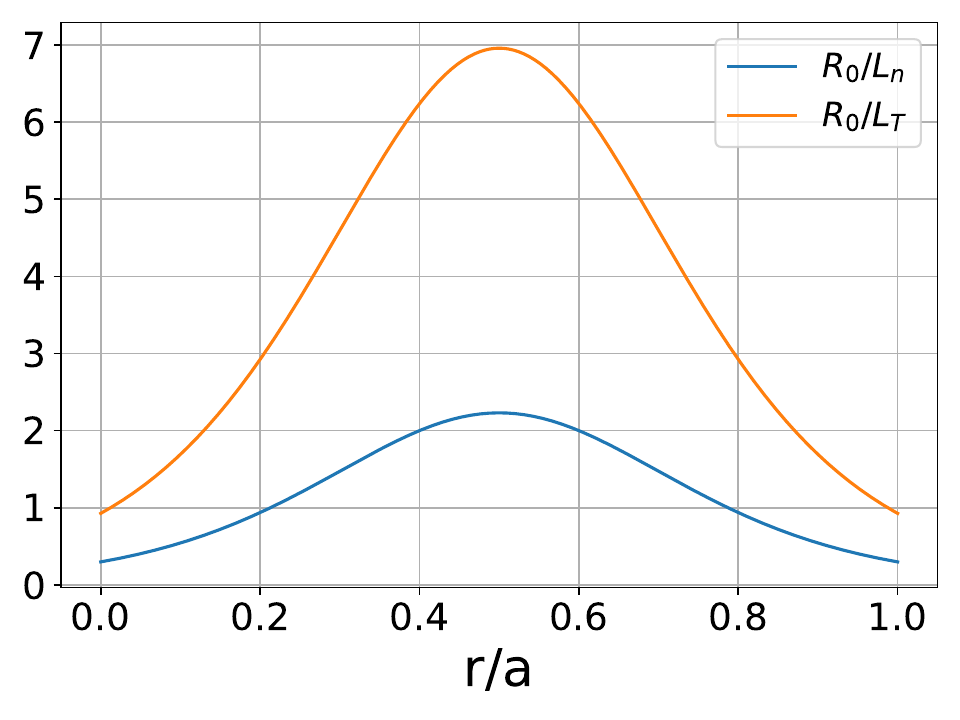}}
   \subfigure[]{
   \label{fig:GrowthRates}
   \includegraphics[width=1.7in]{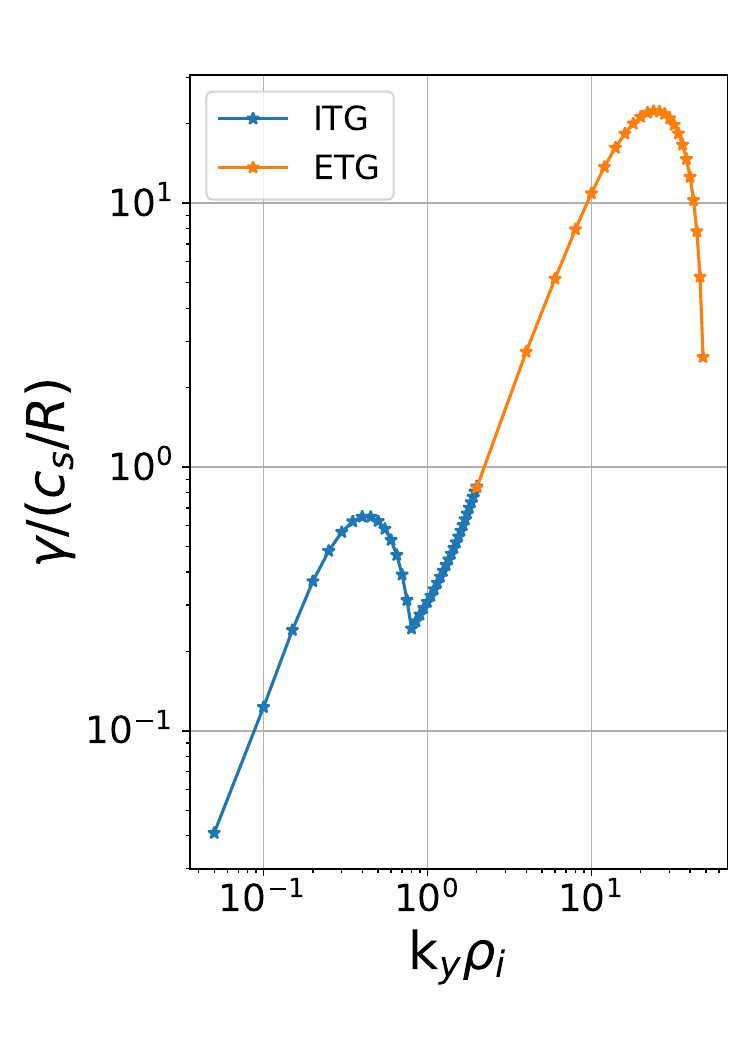}}
   \caption{\label{fig:GeneSetup} (a) Density and temperature profiles and (b) the respective normalized gradient profiles as functions of $r/a$ defined by \eqref{RadProfEqn}
            and \eqref{GradProfEqn}. (c) ITG and ETG growth rates from linear electrostatic GENE simulations at $r/a=0.50$.}
\end{figure*}

Given local electron-scale simulations, a form for the source term can be found by taking the flux-surface-average of \eqref{gkETG}. Due to the periodic boundary conditions
in the flux-tube approximation one finds that the source term must come from the ES $\textbf{E}$x$\textbf{B}$ nonlinearity,
\begin{equation}\label{srcEqn}
   \langle S_{\textrm{ES}} - \frac{1}{B}\delta\textbf{E}_{\textrm{ES}}\times\textbf{b}\cdot\nabla\delta f_{\textrm{ES}}\rangle_{\psi} = 0,
\end{equation}
where $\langle\ldots\rangle_{\psi}$ represents a flux-surface average. Note that the mesoscale average can be used if the turbulence is strongly scale separated, but in
principle the linear terms may not average to zero for intermediate-scale modes. As this source is responsible for maintaining the steady-state transport, it can account
for excess electron thermal transport from ETG modes. 

The second non-zero term is recast as a diffusion operator in real space,
\begin{equation}\label{smDiffEqn}
   \langle\frac{1}{B}\delta\textbf{E}_{\textrm{ES}}\times\textbf{b}\cdot\nabla\delta f'\rangle_{\textrm{m}} \approx \langle D_{\textrm{ES}}\nabla_{\perp}^2\delta f'
   \rangle_{\textrm{m}} = D_{\textrm{ES}}\nabla_{\perp}^2\delta f_{\textrm{IS}}.
\end{equation}
It is expected that the ES $\textbf{E}$x$\textbf{B}$ effects can lead to changes in the IS electron distribution function. This effect can likely be modeled as a
diffusion, $D_{\textrm{ES}}$, acting on the IS distribution function due to the ETG electrostatic potential. Such a model of microturbulence-induced diffusion has
previously been studied in the saturation of energetic-particle-driven modes.\cite{DiffTAE_Lang2011}

Sections \ref{sec:gene} and \ref{sec:gem} address including the ES source term in global ion-scale simulations to capture electron-scale heat transport. Flux-tube ETG
simulations are carried out using GENE at different radial locations. A kinetic source term is constructed using local results at the peak temperature gradient location
and the source term is added to global IS GEM simulations. The ES source term is varied in accordance with various multiscale simulation results, and the result of adding
ES transport to global IS simulation is discussed. Different analytic radial profiles of the ETG heat flux are then compared to results from local ES simulations at
multiple radial locations.

\section{Local Simulations\label{sec:gene}}

Local linear and nonlinear flux-tube simulations of ITG and ETG turbulence were carried out in GENE\cite{Gene1_Jenko2000} to test for a simple scenario with suitable scale
separation. Circular Cyclone Base Case (CBC) parameters\cite{GorlerCBC} are taken for geometric and computational simplicity. Gyrokinetic ions and electrons are used for
simulations at both scales with deuterium chosen as the main ion species, and collisions are included at both scales. As described in Ref. \onlinecite{GorlerCBC}, normalized
radial density and temperature profiles and the associated normalized gradients for both species are given by the following equations:
\begin{equation}\label{RadProfEqn}
   A(r)/A(r_0) = \exp{\left[-\kappa_A w_A\frac{a}{R}\tanh{\left(\frac{r-r_0}{w_A a}\right)}\right]},
\end{equation}
\begin{equation}\label{GradProfEqn}
   R/L_A = -R\partial_r(\ln{A(r)}) = \kappa_A\cosh^{-2}{\left(\frac{r-r_0}{w_A a}\right)},
\end{equation}
for A $\in$ \{$n_0,T_0$\}, $\kappa_A\in\{\kappa_n,\kappa_T\} = \{2.23,6.96\}$ defining the gradient profile peaks, and $w_A = 0.30$ the gradient profile widths.

The density and temperature profiles and their gradients are shown in Fig. \ref{fig:EqProfiles} and \ref{fig:GradProfiles}. The safety factor profile is given by
\begin{equation}\label{SafFacEqn}
   q(r) = 2.52(r/a)^2 - 0.16(r/a) + 0.86,
\end{equation}
with $\hat{s} = \frac{r}{q}\frac{dq}{dr}$ the magnetic shear profile. To retain only the electrostatic instabilities of interest, the plasma beta factor,
$\beta = 8\pi n_{0e}T_{0e}/B_0^2$, with $B_0$ the on-axis magnetic field, is taken to be $\beta = 1e^{-4}$. Multiscale CBC simulations have previously reported
important cross-scale interactions,\cite{MultiscaleCBC_Maeyama2015} and so the simplified circular CBC parameters were chosen to facilitate initial theoretical
investigation. While these parameters are idealized, ETG turbulence profiles can in principle be captured for any experimental scenario by using results from
multiple flux-tube simulations at different radial locations.

\begin{table*}
   \begin{center}
   \caption{Time-averaged heat fluxes for all simulations, in gyroBohm units.}
   \begin{tabularx}{0.8\textwidth}{
        >{\centering\arraybackslash}X
      | >{\centering\arraybackslash}X
      | >{\centering\arraybackslash}X
      | >{\centering\arraybackslash}X
      | >{\centering\arraybackslash}X
      | >{\centering\arraybackslash}X
      | >{\centering\arraybackslash}X}
      \hline
      Run & Code & Scale & $\beta$ & $\gamma_{E'}$ & $Q_i$  & $Q_e$ \\
      \hline
      $\#1$ & GENE & ES          & $1e^{-4}$ & 0   & 0.22   & 30.78 \\
      $\#2$ & GENE & IS          & $1e^{-4}$ & 0   & 171.26 & 50.16 \\
      $\#3$ & GENE & ES          & $1e^{-2}$ & 0.2 & 0.21   & 29.19 \\
      $\#4$ & GENE & IS          & $1e^{-2}$ & 0.2 & 8.82   & 4.79  \\
      \hline
      $\#5$ & GEM  & IS (global) & $1e^{-4}$ & 0   & 74.94 & 14.39 \\
   \end{tabularx}
   \label{tab:nlFlux}
\end{center}
\end{table*}

\begin{figure*}
   \centering
   \subfigure[]{
   \label{fig:FluxSpectra}
   \includegraphics[width=2.3in]{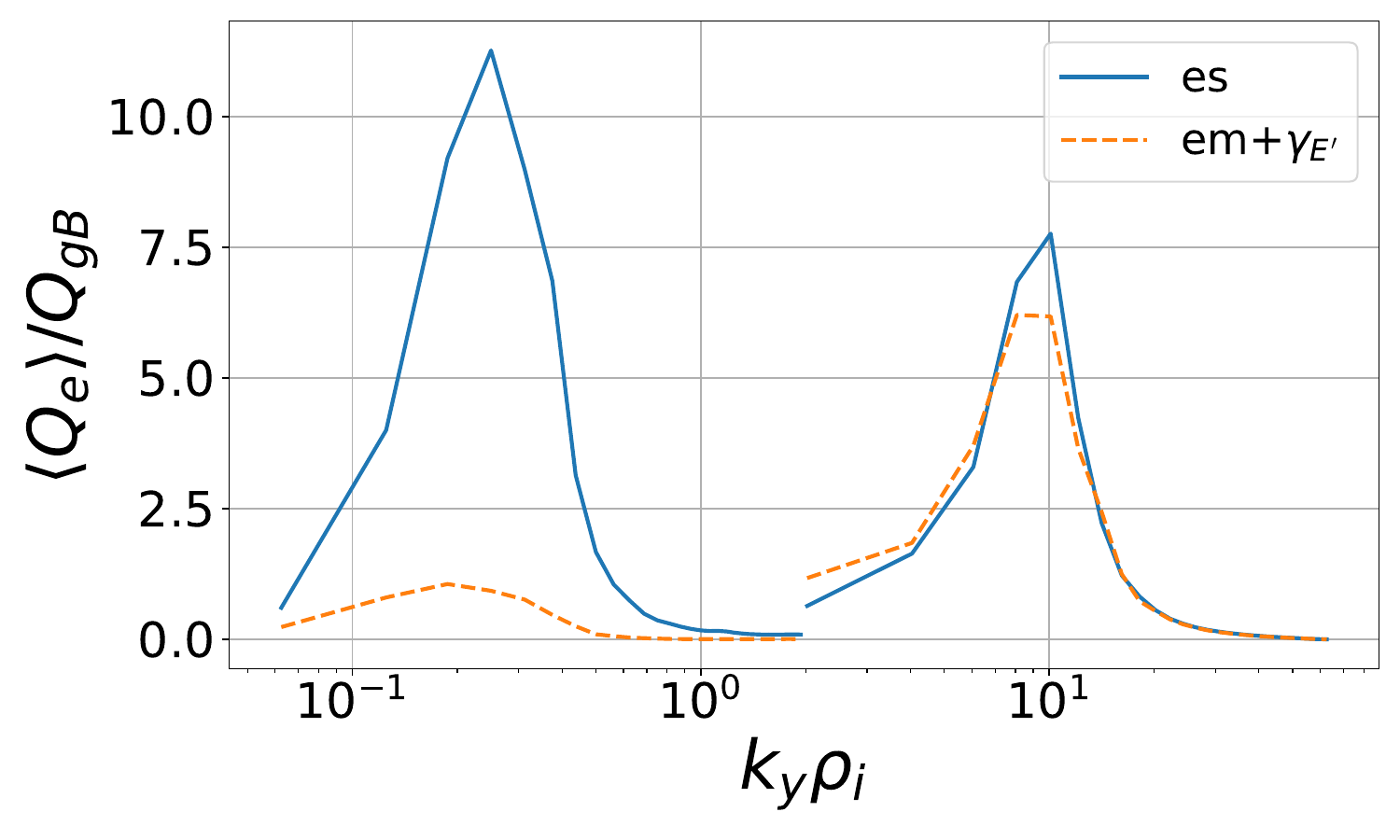}}
   \subfigure[]{
   \label{fig:PhiySpectra}
   \includegraphics[width=2.3in]{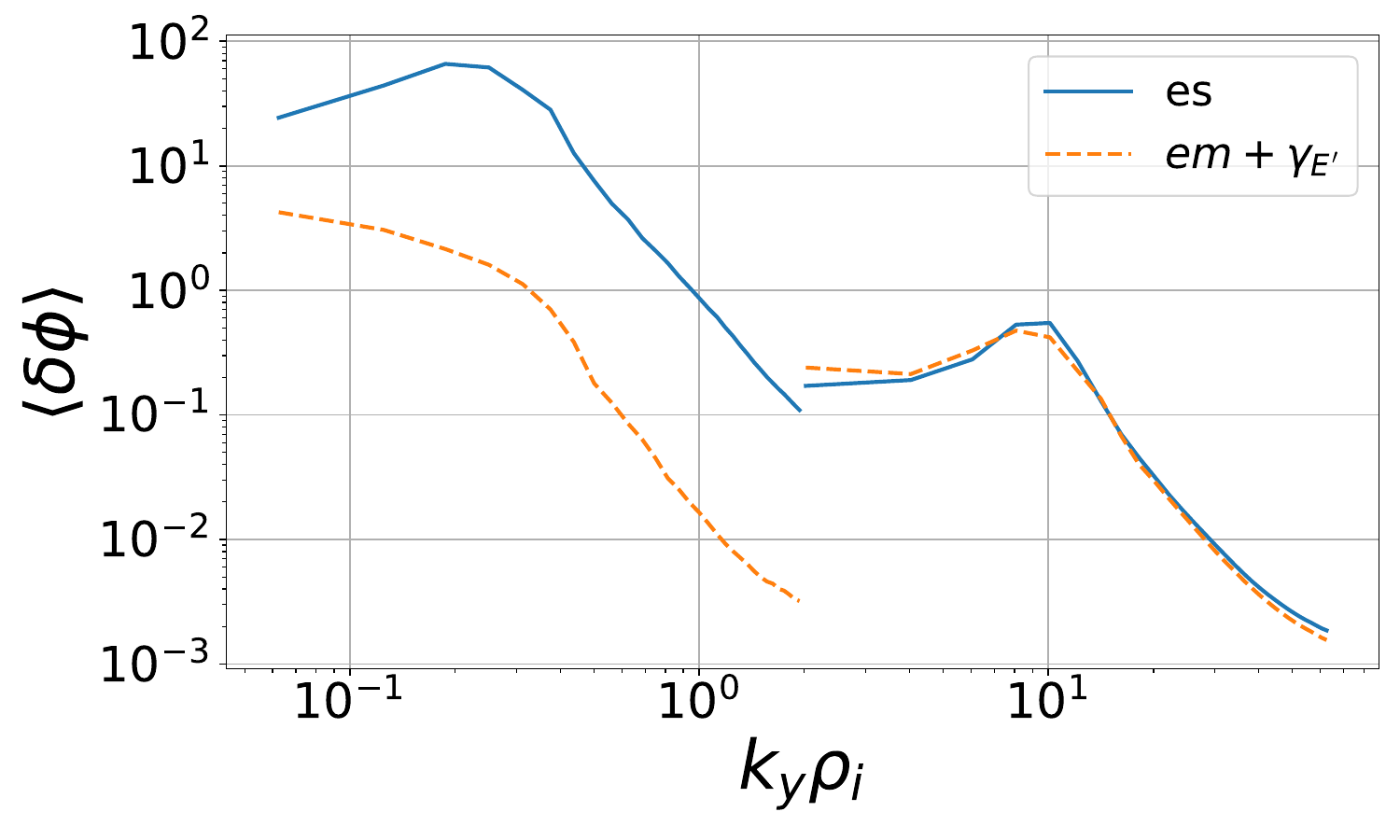}}
   \subfigure[]{
   \label{fig:PhixSpectra}
   \includegraphics[width=2.3in]{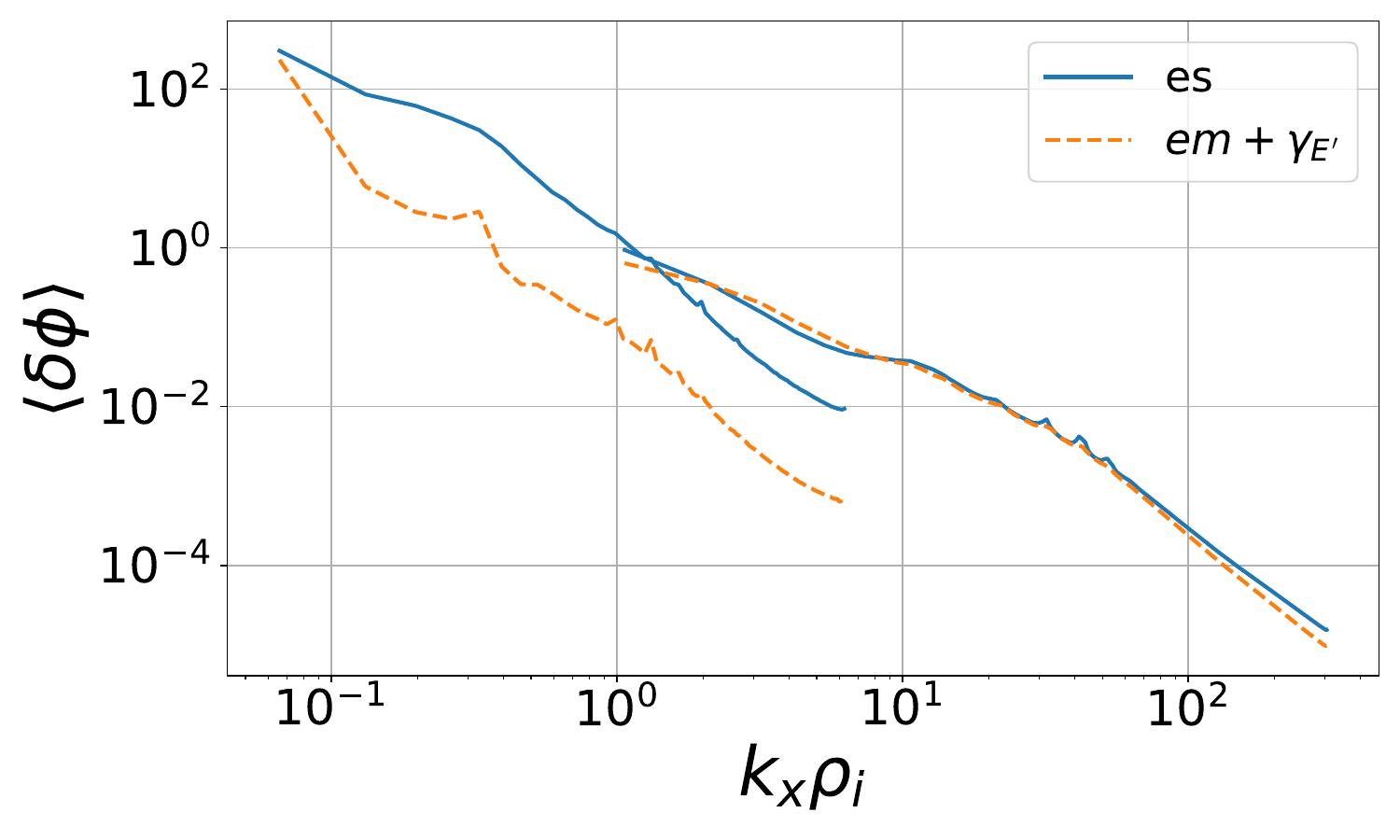}}
   \caption{\label{fig:NonlinearSpectra} Flux spectra for GENE IS and ES flux-tube scenarios. (a) Electron heat flux and (b) electrostatic potential spectra as functions of $k_y$,
            averaged over $k_x$ and $z$. (c) Electrostatic potential spectra as a function of $k_x$, averaged over $k_y$ and $z$. Labels `es' and `em+$\gamma_{E'}$' correspond to
            electrostatic runs and electromagnetic runs with shear flow respectively.}
\end{figure*}

The normalized mode frequencies, $\omega/(c_s/R)$, and growth rates, $\gamma/(c_s/R)$, for the instabilities are calculated using GENE linear electrostatic simulations. Here $c_s
= \sqrt{T_e/m_D}$ is the deuteron sound speed. The linear growth rates at $r/a = 0.50$ are shown in Fig. \ref{fig:GrowthRates} spanning IS and ES scales in $k_y\rho_i$, where
$k_y$ is the wavenumber in the binormal direction. Grid convergence values in $z\times v_{\parallel}\times\mu$ were found by increasing grid resolution until growth rates
were constant to three decimal places. Here, $z$ is the field-line-following coordinate, $v_{\parallel}$ is the particle velocity along a field line, and $\mu = m v_{\perp}^2
/2B$ is the magnetic moment which represents particle velocity perpendicular to a field line. Convergence was checked for the most unstable mode as well as a
longer wavelength mode closest to the peak of the nonlinear flux spectra.

The resulting grid resolutions, $z\times v_{\parallel}\times\mu$, used for linear simulations were $32\times 48\times 16$ at the ion scale and $48\times 48\times 48$ at the
electron scale. For both scales $32$ gridpoints are taken in the local radial coordinate $x$. The IS modes start at $k_y\rho_i = 0.05$ with $40$ modes up to $k_y\rho_i = 2.0$
and the ES modes start at $k_y\rho_i = 2.0$ with $32$ modes up to $k_y\rho_i = 64.0$. A clear scale separation in both frequency and wavenumber can be seen in Fig.
\ref{fig:GrowthRates}. The scale separation is of order $\gamma_{ES}/\gamma_{IS} \sim k_{y,ES}/k_{y,IS} \sim \sqrt{m_D/m_e} \sim 60$, which is expected given the respective
IS and ES orderings of $\rho_i$ and $\rho_e$ in space perpendicular to the field and $a/v_{th,i}$ and $a/v_{th,e}$ in time. With the equilibrium gradients dropping off outward
from the center, growth rates for both scales remained appreciable out to $r/a = 0.35$ and $0.65$ while falling to zero near $r/a = 0.20$ and $0.80$. While the linear
simulation results may include trapped-electron modes (TEMs), primarily in the range $k_y\rho_i\sim 0.50-2.0$, the nonlinear GENE results presented next show little radial
heat flux or electrostatic potential in $k_y$ at this intermediate range.

The nonlinear simulations are discussed here in detail for both scales at $r/a = 0.50$. The nonlinear $z\times v_{\parallel}\times\mu$ grid resolution was reduced to lower
values for which growth rates were still converged to within $\pm 1\%$. The new grid resolutions are $16\times 32\times 8$ and $32\times 32\times 16$ for the IS and ES
scales. Nonlinear IS simulations were also tested with an increased resolution of $32\times48\times16$, while for ES simulations the resolution in each of the three
dimensions was increased by a factor of $2$ respectively to test for convergence. The ITG runs include $32$ modes ranging from $k_y\rho_i = 0.0625$ to $2.0$, and ETG runs
include $32$ modes ranging from $k_y\rho_i = 2.0$ to $64.0$. The perpendicular domain sizes, $L_x\times L_y$, are $96\rho_i\times 100\rho_i$ and $360\rho_e\times 190\rho_e$,
with the $L_x : L_y$ ratio increased for ETG runs to allow for possible generation of intermediate-scale zonal flows. The radial grid resolutions are $\Delta x = 0.50\rho_i$
and $\Delta x = 0.62\rho_e$ respectively. The perpendicular domain for ETG runs was tested by separately increasing $L_x$ and $L_y$ by a factor of $2$ each.

Ion and electron heat fluxes, $Q_i$ and $Q_e$, are listed in Table \ref{tab:nlFlux} for all ES and IS nonlinear runs. Heat fluxes are normalized using the gyroBohm value
$Q_{gB} = n_e c_s T_e(\rho_D/R)^2$, with $\rho_D$ the deuteron gyroradius. Runs $\#1$ and $\#2$ correspond to ES and IS simulations without background shear flow or
electromagnetic effects. Runs $\#3$ (ES) and $\#4$ (IS) include these effects to regulate ITG turbulence\cite{emITG_Kim93,erITG_Kinsey05} in the unstable CBC scenario,
providing a comparison to multiscale scenarios in which ITG turbulence is marginal. This results in a $95\%$ reduction in $Q_i$. By comparison, simulations with only shear
flow or finite-beta effects show reductions of approximately $37.5\%$ and $85\%$ respectively. The value of $\beta$ is derived directly from the equilibrium profiles.
The choice of shearing rate, $\gamma_{E'} \approx \gamma_{\textrm{ITG}}^{\max}/2= 0.2$, comes from comparing the most unstable ITG growth rate in the electromagnetic case
to the maximum shearing rate within experimental uncertainty for the DIII-D IBS values from Ref. \onlinecite{MultiscaleIBS_Holland2017}. The heat fluxes are further compared
to results from the global ion-scale GEM simulation (run $\#5$) which is discussed further in section \ref{sec:gem}. The GEM simulation ranges from $r/a = 0.20$ to $0.80$,
and the heat flux is calculated by averaging over a centered domain spanning from $r/a=0.35$ to $0.65$.

These scenarios provide variable ratios of IS and ES turbulence for comparison with multiscale simulations results. In runs $\#1$ and $\#5$, the ratio of ES to total electron
heat transport falls within the approximate range of $1/2-2/3$ seen in multiscale scenarios with appreciable ES effects.\cite{Multiscale2_Howard16,MultiscaleIBS_Holland2017}
Finite-beta and shear flow effects are not included directly in GEM as the CBC scenario with electromagnetic effects is often numerically challenging.\cite{WaltzEM} Instead,
ES effects are magnified by a factor of $3$ when included in GEM in section \ref{sec:gem}. This increased factor is used to match the ratio of electron heat flux in runs
$\#3$ and $\#4$ when ITG is regulated. The magnified ES scenario represents cases of marginal ITG turbulence in multiscale scenarios, where the ratio of total $Q_e : Q_i$ 
can range from $1$-$3$.\cite{Multiscale2_Howard16,MultiscaleIBS_Holland2017,MultiscaleJet_Bonanomi2018}

For Runs $\#1-4$, the spatial averages of the heat flux spectra are shown as functions of $k_y$ in Fig. \ref{fig:FluxSpectra}. The spatial averages of the electrostatic
potential spectra are further shown as functions of $k_y$ in Fig. \ref{fig:PhiySpectra} and $k_x$ in Fig. \ref{fig:PhixSpectra}. Electrostatic simulation results are
labeled `es', while the electromagnetic results with shear flow are labeled `em+$\gamma_{E'}$'. The difference in ITG turbulence levels between both cases is clear. A
strong spatial separation of scales exists in $k_y$ for both the heat flux and potential spectra. However, the $k_x$ potential spectrum is continuous, due to the
generation of intermediate-scale zonal flows at the electron scale. Such zonal flows have been reported in multiscale simulations with marginal levels of ITG turbulence,
where they may further suppress ion-scale fluctuations.\cite{MultiscaleIBS_Holland2017} Capturing these effects would require breaking scale-separation assumptions and
this is briefly mentioned further in section \ref{sec:disc}.

The choice of mesoscale values in space perpendicular to the field and in time, $l_{\textrm{m}}$ and $\tau_{\textrm{m}}$, is based on the peaks of the nonlinear $k_y$
spectra, rather than peaks of the linear growth rate. For the electrostatic cases, the peaks occur at $k_y\rho_i$ = $0.25$ and $10.1$ for the IS and ES scales, giving
length scales of $25\rho_i$ and $0.62\rho_i$. The linear mode frequencies for these wavenumbers are $\omega/(c_s/R)$ = $0.586$ and $17.1$, giving time scales of $10.72R/c_s$
and $0.367R/c_s$. The separation of scales is then approximately $40x$ in space and $29x$ in time, compared to the theoretical estimate of $60x$ seen in the linear
simulation results. The mesoscale length is taken at $k_y\rho_i$ = 2.0, where the linear mode frequency is $\omega/(c_s/R) = 2.8$, giving $l_{\textrm{m}}=3.14\rho_i$ and
$\tau_{\textrm{m}}=2.24R/c_s$. This choice of $k_y\rho_i$ then allows for using a perpendicular flux-surface-average in lieu of a true intermediate-scale averaging since
this represents the extent of the $y$-domain. Since only $k_y$ modes contribute to the radial transport, the issue of scale separation in $x$ is ignored for now, and the
whole radial domain is averaged over.

Finally, multiple ES simulations were carried out at various radial locations. The global radial electron heat flux profile is shown later in Fig. \ref{fig:QeRadComp} in
comparison to the various global theoretical models considered in section \ref{sec:gem}. Necessary simulation parameters were updated accordingly using \eqref{RadProfEqn}-
\eqref{SafFacEqn}. The radial grid resolution was also reduced by a factor of $2$ such that $\Delta x = 1.24\rho_e$ to decrease computational cost. The new value of
$Q_e/Q_{gB}$ at $r/a=0.50$ for the updated case was $28.24$, similar to the original $30.78$. Local IS simulations were not performed at multiple radii.

\section{Global Ion-Scale Simulations\label{sec:gem}}

\begin{figure}
   \includegraphics[width=3in]{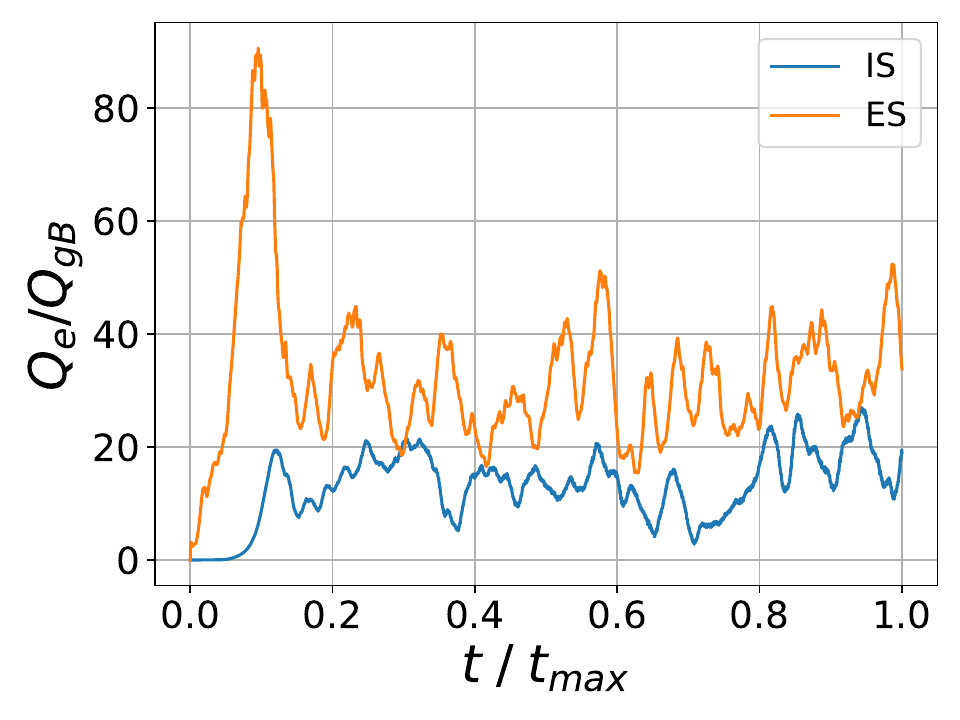}
   \caption{\label{fig:QeComps}Comparison of surface-averaged heat fluxes over time for local GENE ES and global GEM IS cases (Runs $\#1$ and $\#5$). The final times,
            $t_{\max}$, are $27.50R/c_s$ for the ES run and $147.85R/c_s$ for the IS run.}
\end{figure}

Nonlinear nonlocal electrostatic gyrokinetic simulations are carried out in the $\delta f$ particle-in-cell code GEM.\cite{Gem1_Chen2003,Gem2_Chen2007} The grid resolution
is $256\times128\times64$ in the radial, binormal, and parallel directions. $16$ ions and $32$ electrons per cell are used. The time step is $\Omega_p\Delta t = 1$, with
$\Omega_p$ the proton cyclotron frequency, and the radial domain is $0.20 \le r/a \le 0.80$. Drift-kinetic electrons are included using a split-weight scheme.\cite{Gem2_Chen2007}
Each particle is given a weight defined by $w=\delta f/f_M$, where $f_M$ is the Maxwellian distribution. Electron weights are evolved in time according to the equation
\begin{equation}
\begin{aligned}
   \dot{w_e} = &-\left[\textbf{v}_E\cdot\frac{\nabla f_M}{f_M} - \frac{e}{T_{0e}}(\textbf{v}_G\cdot\nabla\delta\phi)\right. \\
   &+ \left.\frac{e}{T_{0e}}(\partial_t\delta\phi(\textbf{x}) + \textbf{v}_G\cdot\nabla\delta\phi\vert_{\textbf{x}}) -\frac{S}{f_M}\right]\frac{f_M}{g_0},
\end{aligned}
\end{equation}
with $g_0$ the marker particle distribution, $\textbf{v}_E$ the $\textbf{E}$x$\textbf{B}$ drift, $\textbf{v}_G$ the guiding-center drift, and $\textbf{x}$ the particle
location. $S$ represents a numerical heat source which maintains the steady-state transport.\cite{Gem3_Chen2023} The heat flux in GEM is averaged over a toroidal annulus
extending radially from $r/a = 0.35$ to $0.65$. Fig. \ref{fig:QeComps} compares the electron heat fluxes over time for the local ES ($\#1$) and global IS ($\#5$)
simulations.

The changes to GEM discussed here account for excess electron heat transport from the electron scale by including the new source term in \eqref{gkITG}. This
term is responsible for radial $\textbf{E}$x$\textbf{B}$ transport caused by electron-scale turbulence in ion-scale simulations, and can be broken up into total
divergence and compressible flow terms
\begin{equation}\label{srcFullETG}
\begin{aligned}
   \langle S_{\textrm{ES}}\rangle_{\textrm{m}} &= \langle\frac{1}{B}\delta\textbf{E}_{\textrm{ES}}\times\textbf{b}\cdot\nabla\delta f_{\textrm{ES}}\rangle_{\textrm{m}} \\
   &= \langle\nabla\cdot(\textbf{v}_{E,\textrm{ES}}\delta f_{\textrm{ES}}) - \delta f_{\textrm{ES}}(\nabla\cdot\textbf{v}_{E,\textrm{ES}})\rangle_{\textrm{m}} \\
   &= \langle\nabla_r\cdot(\textbf{v}_{E,\textrm{ES}}\delta f_{\textrm{ES}})\rangle_{\textrm{m}}
    + \cancel{\langle\nabla_{x}\cdot(\textbf{v}_{E,\textrm{ES}}\delta f_{\textrm{ES}})\rangle_{\textrm{m}}} \\
   &\;\;\;\;\;- \langle\delta f_{\textrm{ES}}(\nabla\cdot\textbf{v}_{E,\textrm{ES}})\rangle_{\textrm{m}} \\
   &= \nabla_r\cdot\langle\textbf{v}_{E,\textrm{ES}}\delta f_{\textrm{ES}}\rangle_{\textrm{m}}
    - \langle\delta f_{\textrm{ES}}(\nabla\cdot\textbf{v}_{E,\textrm{ES}})\rangle_{\textrm{m}}.
\end{aligned}
\end{equation}
Here, $v_E$ is the $\textbf{E}$x$\textbf{B}$ drift velocity which, in the case of electrostatic waves in toroidal geometry, varies as $\nabla\cdot v_E \sim v_E/R$.
The global radial divergence is pulled out of the local intermediate-scale spatial average, and, due to the periodic boundary conditions of the flux-tube
approximation, the total divergence is zero for ES fluctuations.

A crude assumption can be made by focusing on the effects of ETG heat flux to assume a global radial variation $\nabla_r\cdot Q_e \sim Q_e/L_T$, so that the
radial variation of both terms can be compared:
\begin{equation}\label{srcTermComps}
   \frac{\nabla\cdot\langle(\delta f_{\textrm{ES}}\textbf{v}_{E,\textrm{ES}})\rangle_{\textrm{m}}}
        {\langle\delta f_{\textrm{ES}}\nabla\cdot\textbf{v}_{E,\textrm{ES}}\rangle_{\textrm{m}}} \sim 
   \frac{\langle\delta f_{\textrm{ES}}\textbf{v}_{\textrm{ES}}\rangle_{\textrm{m}}/L_T}
        {\langle\delta f_{\textrm{ES}}\textbf{v}_{\textrm{ES}}\rangle_{\textrm{m}}/R} \sim R/L_T.
\end{equation}
The compressibility term can then be ignored, as $R/L_T$ ranges from $3$ to $7$ when $r\;\in\; [0.2,0.8]$ for the CBC profiles shown in Fig. \ref{fig:GradProfiles}.
However, this assumption is only reasonable when taking the second moment of these terms in velocity space. This allows for focusing on ETG heat transport, but at the
loss of compressible effects regarding other moments as the ETG particle and momentum flux are considered negligible. Any possible contribution from the Reynolds
stress of ETG turbulence is also lost. Notably, inclusion of the ETG Reynolds stress can result in an effective dissipation of TEM modes in good agreement with
multiscale scenarios.\cite{MultiscaleTEM_Watanabe2023}

Due to the choice of perpendicular spatial mesoscale, $l_{\textrm{m}}$, discussed in section \ref{sec:gene}, the perpendicular spatial mesoscale average is replaced
with a flux-surface average. The $z$-average has been retained for simplicity and is defined as
\begin{equation}
\langle\ldots\rangle_z = \frac{\int (\ldots)J(z)dz}{\int J(z)dz}.
\end{equation}
In the general subgrid model formulation, any dependence on $z$ must be preserved as the source can depend on the poloidal angle. The flux-surface average is only
employed here to more readily test adding the ES heat transport to GEM. The final ES source then becomes
\begin{equation}\label{SrcETG}
   \langle S_{ES}\rangle_{\textrm{m}} \approx \nabla_r\cdot\langle\langle\textbf{v}_{E,\textrm{ES}}\delta f_{\textrm{ES}}\rangle_{\tau_{\textrm{m}}}\rangle_{\psi}
   = \nabla_r\cdot\hat{\Gamma}_{\textrm{ETG}}(r,v_{\parallel},\mu).
\end{equation}

In this context, $\hat{\Gamma}$ is used instead of $\Gamma$ to indicate that it is not a particle flux, but rather a kinetic form that has been both flux-surface-averaged
and time-averaged. Hereafter, $\hat{\Gamma}$ will be referred to as the kinetic flux density. It is important to note that the $z$-averaging process does not commute with
velocity-space integration, and the kinetic flux density is further normalized to yield the correct heat flux upon integration with factor $\frac{1}{2}mv^2$. Consequently,
the particle flux differs from the exact value. However, because the ion response to ETG turbulence is typically adiabatic, the particle flux is negligible, and the focus
remains on the heat flux.

A simple initial form for $\hat{\Gamma}$ can be constructed by using the diffusion coefficient at the peak temperature gradient location. All radial variation is
retained in the equilibrium temperature profile to obtain a Fick's law diffusive model
\begin{equation}\label{SrcETG2}
\begin{aligned}
   \langle S_{ES}\rangle_{\textrm{m}} &= \nabla_r\cdot\hat{\Gamma}_{\textrm{ETG}}(r,v_{\parallel},\mu) \\
   &= \nabla_r\cdot(-\hat{D}^*_{0,\textrm{ETG}}(v_{\parallel},\mu)\nabla_rT_{0e}(r)) \\
   &= -\hat{D}^*_{0,\textrm{ETG}}(v_{\parallel},\mu)\nabla_r^2T_{0e}(r).
\end{aligned}
\end{equation}
The subscript ETG is chosen here to not conflict with $D_{ES}$ of \eqref{smDiffEqn}, and the temperature gradient is used to recover an appropriate ETG heat flux
when the second moment is taken. This pseudo ETG diffusion coefficient, $\hat{D}^*_{0,\textrm{ETG}}$, is defined by dividing the peak kinetic flux density by
the peak temperature gradient and density at r$_0 = 0.50a$,
\begin{equation}\label{D0_ETG}
   \hat{D}^*_{0,\textrm{ETG}} = -\hat{\Gamma}_{0,\textrm{ETG}}/(n_{0e}\nabla_rT_{0e})\vert_{r_0}.
\end{equation}
$D^*$ has been used to differentiate from the actual diffusion coefficient. This model then allows for correctly capturing the heat diffusivity when taking the
second moment while maintaining negligible particle transport. The radial Laplacian is taken in cylindrical coordinates, giving the normalized value
\begin{equation}\label{grad2T0}
\begin{aligned}
   &-R^2\frac{\nabla_r^2T_{0e}}{T_{0e}} = -R^2\frac{\frac{1}{r}\partial_r(r\nabla_rT_{0e})}{T_{0e}} =\\
   &-\left[\kappa_T^2\textrm{sech}^2(\frac{r-r_0}{w_Ta}) + 2\frac{\kappa_TR}{w_Ta}\tanh(\frac{r-r_0}{w_Ta})\right]\\
   &\;\;\;\;\;\;\;\;\;\;\;\;\;\;\;\;\;\;\;\;\;\;\;\;\;\;\;\;\;\;\;\;\times\textrm{sech}^2(\frac{r-r_0}{w_Ta}) + \frac{R}{r}\frac{R}{L_T}.
\end{aligned}
\end{equation}
The negative sign is added for consistency with \eqref{GradProfEqn}.

\begin{figure}
   \centering
   \includegraphics[width=3.4in]{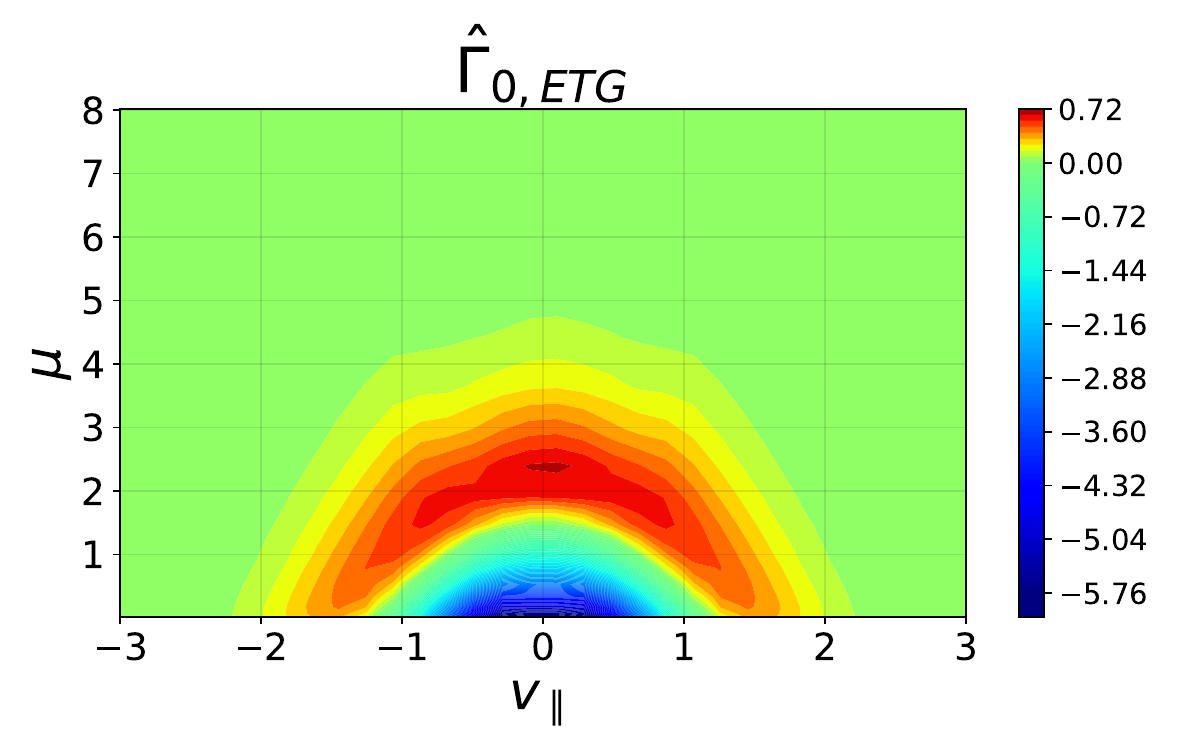}
   \caption{\label{fig:MesoscaleContour}Mesoscale and $z$ average of kinetic flux density taken from GENE ES run $\#1$ at $r/a=0.50$.}
\end{figure}

$\hat{\Gamma}_{0,\textrm{ETG}}(v_{\parallel},\mu)$ is shown in Fig. \ref{fig:MesoscaleContour} above, where the mesoscale time average is taken over $t/t_{\max} = 0.458-
0.540$ in Fig. \ref{fig:QeComps}. This corresponds to the same heat flux listed in Table \ref{tab:nlFlux} which was averaged over the full nonlinear phase, $t/t_{\max} =
0.182-1.0$. The global radial variation is added in accordance with \eqref{SrcETG2}, and the source term converted to GEM normalizations as described in Appendix
\ref{sec:norm}. The ES source term is then included in GEM according to the updated weight and vorticity equations:
\begin{equation}\label{weightEqn}
   \dot{w}_{e} = \dot{w}_{e,\textrm{GEM}} - \frac{\langle S_{\textrm{ES}}\rangle_{\textrm{m}}}{g_0},
\end{equation}
\begin{equation}\label{vortEqn}
   -n_p(\dot{\delta\phi}) = q\partial_t\langle\delta n_i\rangle_{\alpha} - e\partial_t\delta n_e,
\end{equation}
where changes to $\partial_t\delta n_e$ in GEM must reflect changes in the density due to the new source in the electron weights.

Simple diagnostic equations can be used to understand the effect of the new ES term. By focusing on the change in time of the IS distribution function in \eqref{gkITG}
due only to the ES source term, \eqref{SrcETG2}, one can take moments to find
\begin{equation}\label{srcDensTemp}
\begin{aligned}
   \partial_t\delta n_{\textrm{IS}} &= D^*_{0,\textrm{ETG}}\nabla_r^2T_{0e} \approx 0, \\
   \partial_t\delta T_{\textrm{IS}} &= \frac{2}{3}\frac{1}{n_{0e}}\chi_{0,\textrm{ETG}}\nabla_r^2T_{0e},
\end{aligned}
\end{equation}
where $\chi_{0,\textrm{ETG}} = -Q_{0,\textrm{ETG}}/(n_{0e}\nabla_rT_{0e})\vert_{r_0}$ is the ETG heat diffusivity at $r/a=0.50$. Pressure isotropy has been assumed as
there as there is no rotational flow, and the temperature equation is found by linearizing the standard equation of state, $p=nT$, and solving for the change in the
pressure perturbation
\begin{equation}\label{srcPressure}
\begin{aligned}
   \partial_t{\delta p_{\textrm{IS}}} &= \cancel{\partial_t(\delta n_{\textrm{IS}}T_{0e})} + \partial_t(\delta T_{\textrm{IS}}n_{0e}) \\
   &= -\int\frac{2}{3}(\frac{1}{2}mv^2)\langle S_{\textrm{ES}}\rangle_{\textrm{m}}d^3v.
\end{aligned}
\end{equation}

\begin{figure}
   \centering
   \includegraphics[width=3in]{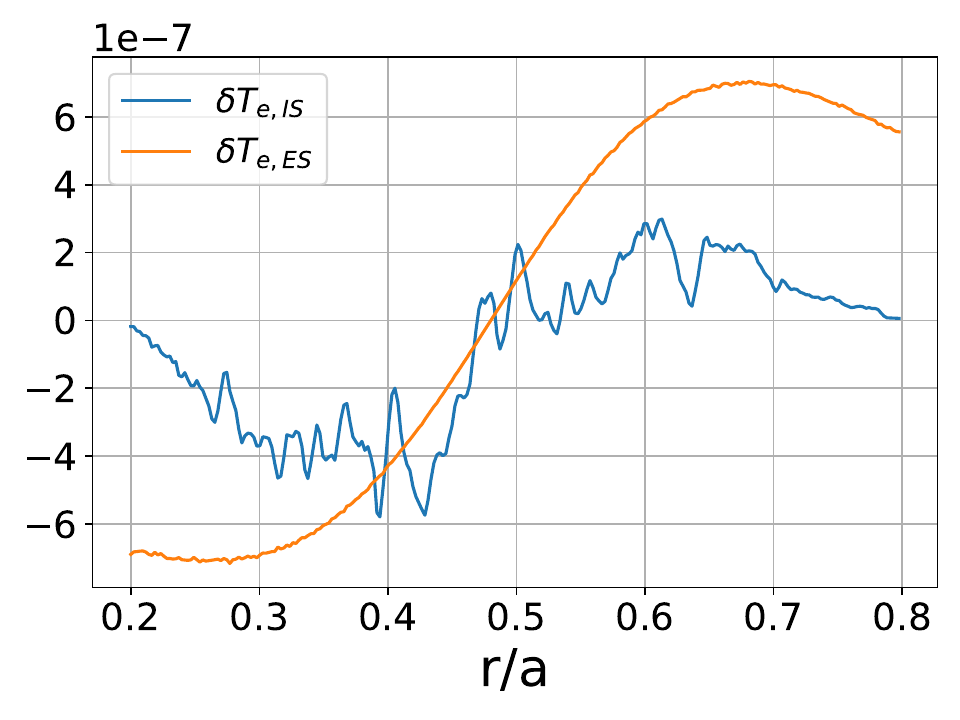}
   \caption{\label{fig:T1eComp}Comparison of the electron temperature perturbations generated by ITG (IS) and ETG (ES) turbulence in GEM.}
\end{figure}

Temperature perturbations are calculated by integrating over particle trajectories in time in GEM. The IS and ES contributions are separated and compared in Fig.
\ref{fig:T1eComp}. The IS perturbation is calculated by averaging over all time in the standard GEM case (run $\#5$) without any subgrid contribution included. The
ES perturbation is calculated by integrating just the $\langle S_{\textrm{ES}}\rangle_{\textrm{m}}$ term over one time step as the subgrid term is constant over time.
The effect of these temperature perturbations is to increase $T_e$ when $r/a \gtrsim 0.50$ and decrease $T_e$ when $r/a \lesssim 0.50$, thus flattening the electron
temperature profile and reducing the possible ITG transport.

\begin{figure}
   \centering
   \subfigure[]{
   \label{fig:QeGemComps}
   \includegraphics[width=3.2in]{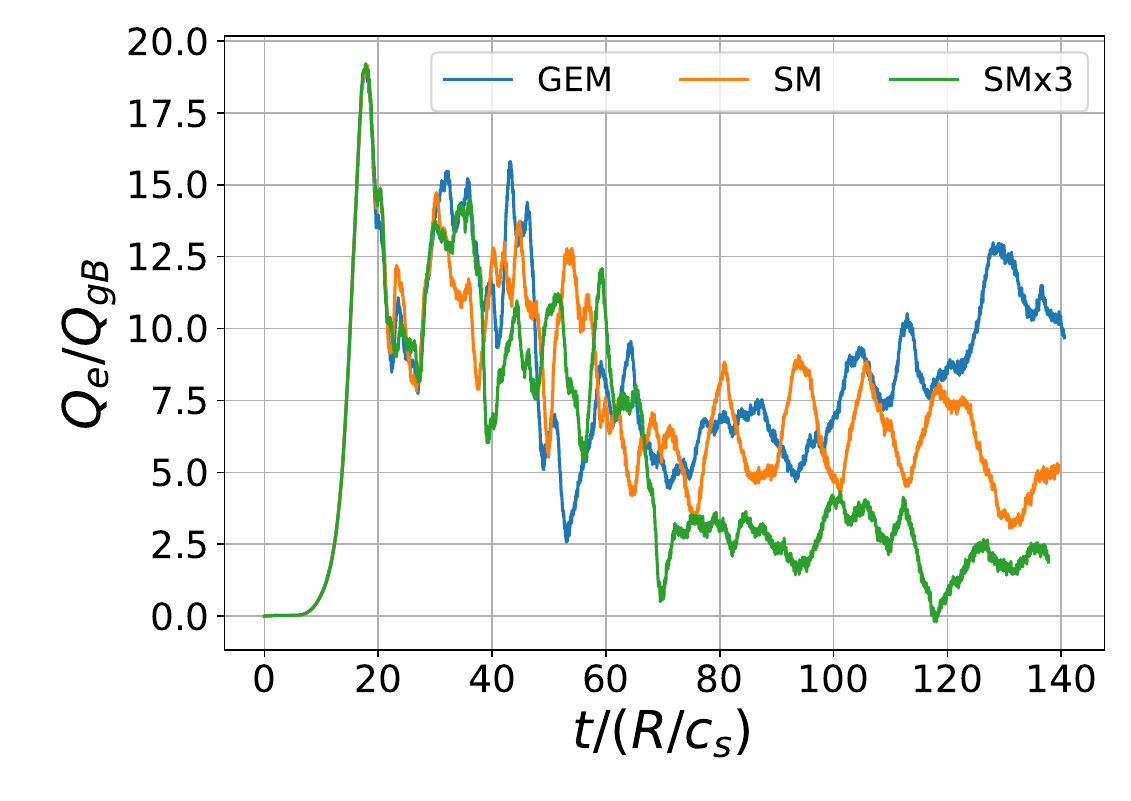}}
   \quad
   \subfigure[]{
   \label{fig:T0eGemComps}
   \includegraphics[width=3.2in]{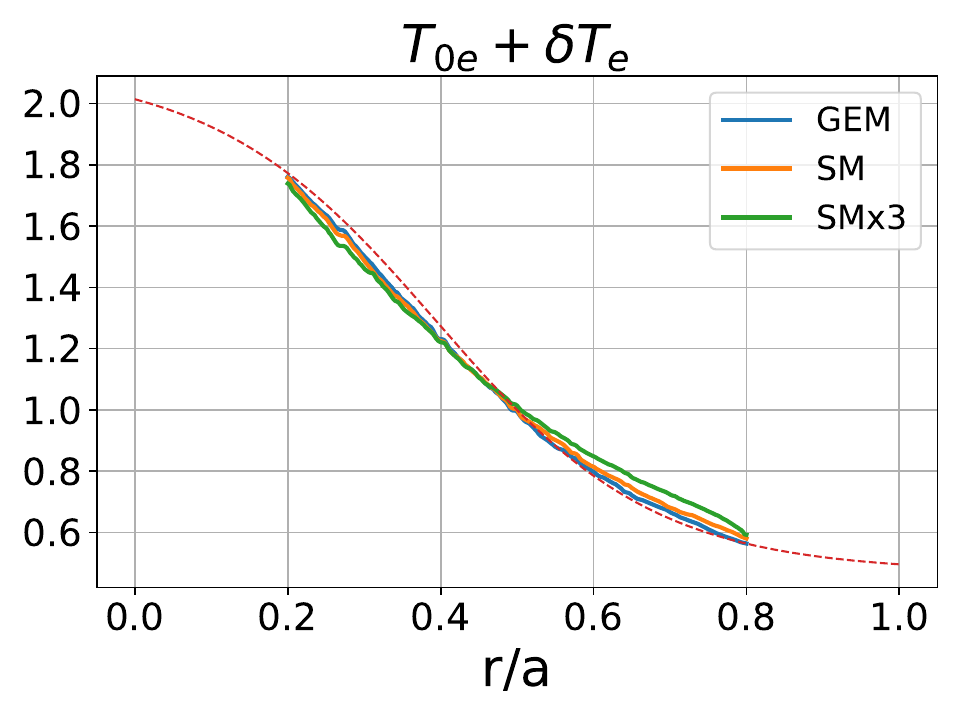}}
   \caption{\label{fig:GemResults}Comparison of (a) electron heat flux evolution averaged over a toroidal annulus from $r/a = 0.35$ to $0.65$ and (b) electron temperature profile flattening for
            three different GEM scenarios at $t=129.75R/c_s$. `GEM',`SM', and `SMx3' correspond to standard GEM, GEM with subgrid model, and GEM with enhanced (3x factor)
            subgrid model runs. The dashed red line corresponds to the initial electron temperature profile.}
\end{figure}

To observe the change in electron heat flux over time, the original heat source in GEM was removed and three simulations were run: one with no subgrid term, one with the
subgrid term included, and one with the subgrid term increased in magnitude. As discussed previously in section \ref{sec:gene}, the increased magnitude ($3x$) comes from
scaling the subgrid term in GEM to match the electron heat flux ratio, $Q_{e,ES}:Q_{e,IS}$, observed in runs $\#3$ and $\#4$. Runs $\#3$ and $\#4$ include
electromagnetic and shear flow effects which suppress ion-scale turbulence, so that this adjusted ratio in GEM represents multiscale scenarios with marginal ITG turbulence.

\begin{figure*}
   \centering
   \subfigure[]{
   \label{fig:LinearContour}
   \includegraphics[width=2.3in]{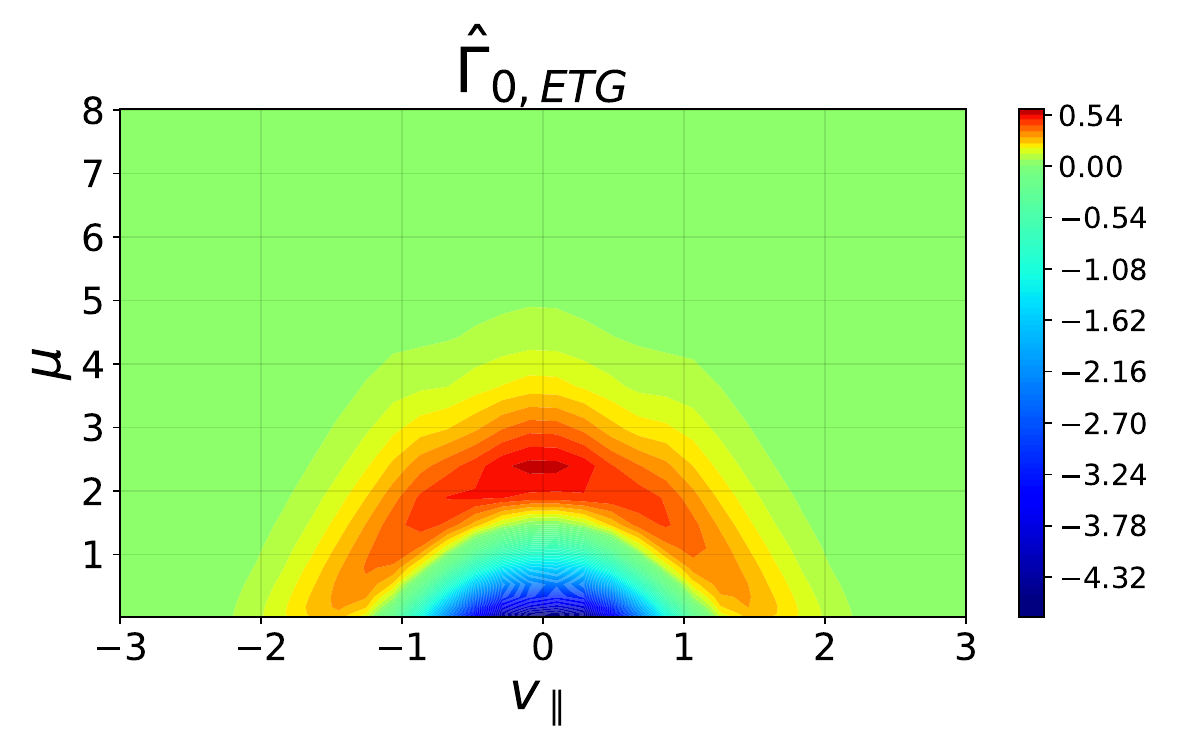}}
   \subfigure[]{
   \label{fig:NonlinearContour}
   \includegraphics[width=2.3in]{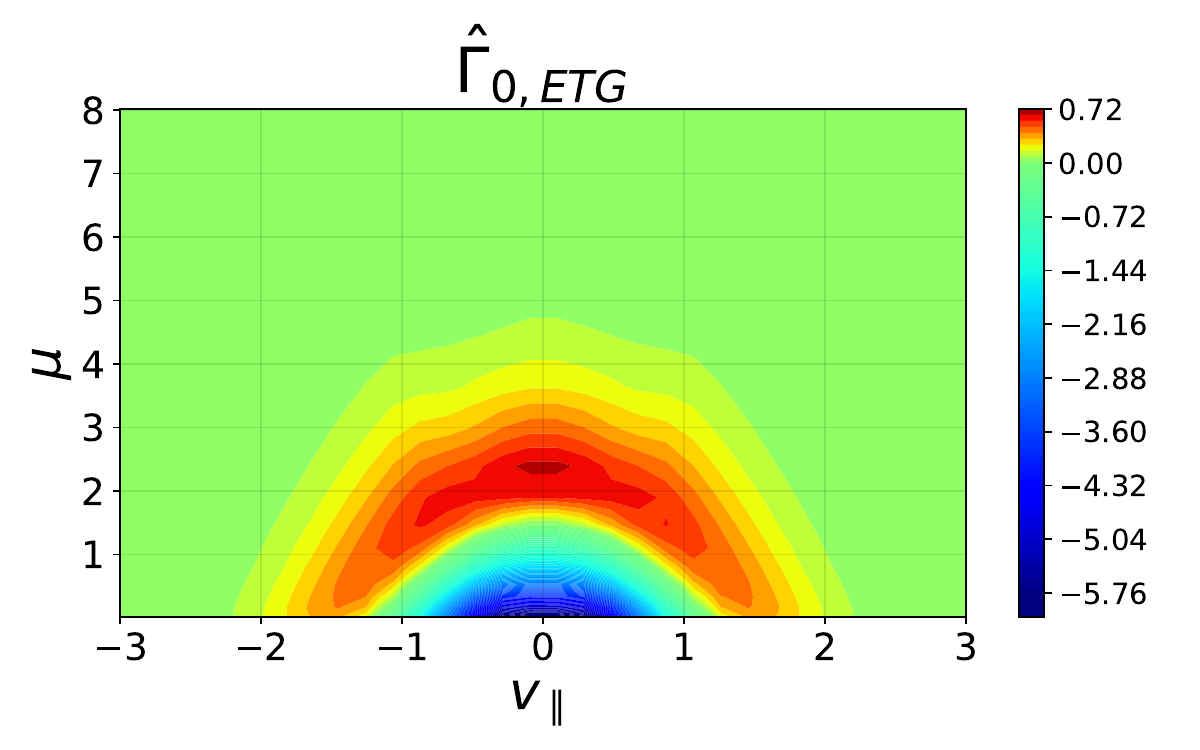}}
   \subfigure[]{
   \label{fig:QuasilinearContour}
   \includegraphics[width=2.3in]{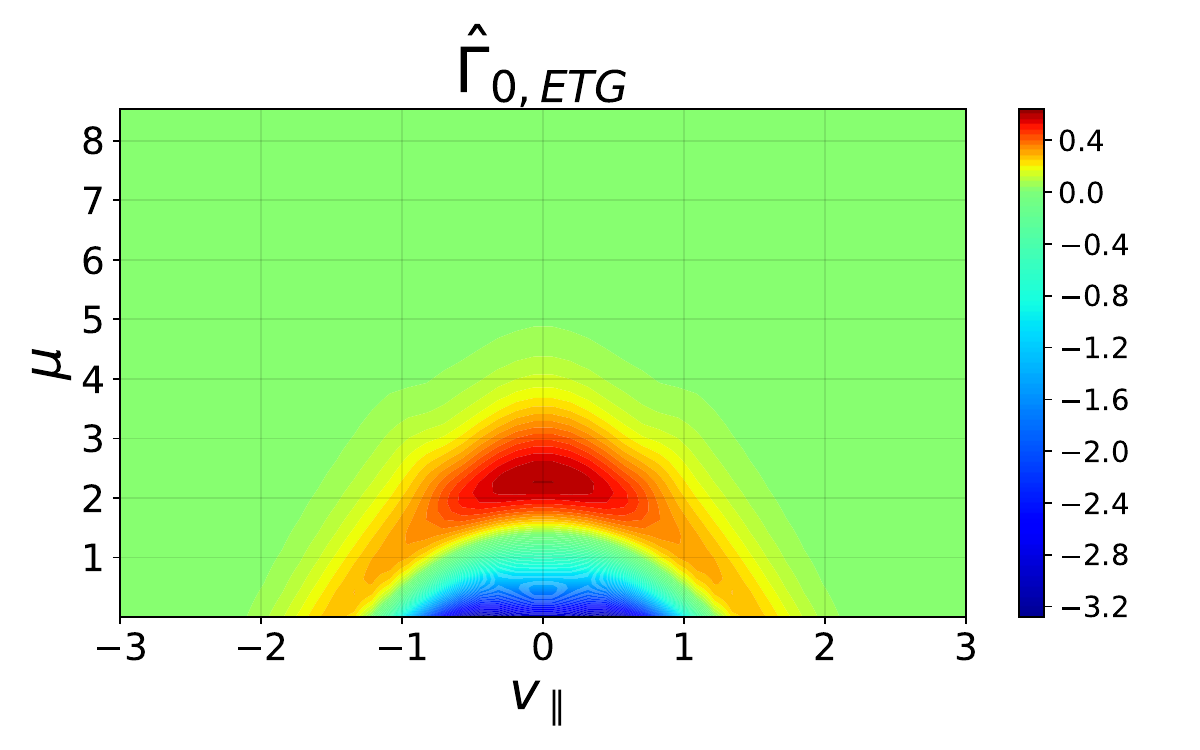}}
   \caption{\label{fig:FluxContours}Kinetic flux density averaged in time over (a) the linear phase, $t/t_{\max} = 0.007-0.072$, and (b) the nonlinear saturated phase,
            $t/t_{\max} = 0.182-0.982$, of run $\#1$ shown in Fig. \ref{fig:QeComps}. (c) Quasilinear model of kinetic flux density using linear GENE simulations. All
            plots at $r/a=0.50$.}
\end{figure*}

Fig. \ref{fig:QeGemComps} compares the difference in electron heat flux over time for the three scenarios. Heat fluxes are calculated by averaging over a toroidal
annulus with radial extent $r/a =0.35$ to $0.65$. Fig. \ref{fig:T0eGemComps} compares the corresponding temperature profile flattening, with the time-averaged
perturbations calculated from times $t = 0 - 129.75R/c_s$. As the subgrid contribution is increased, the electron temperature profile flattens more quickly and the 
ITG mode drives less electron thermal transport. Although the source term directly affects the electron distribution function, the ITG turbulence was unaffected and
no meaningful differences in the ion particle and heat fluxes or electron particle flux were observed.

Lastly, the validity of the Fick's law diffusive model is discussed. A key limitation of this model is that all radial variation is attributed to the temperature gradient.
Since the electron temperature gradient remains non-zero at the ends of the domain, the source term does not vanish. However, as discussed in section \ref{sec:gene}, the
ETG modes are stabilized near the boundaries and there should be no ES heat transport. This issue is evident in Fig. \ref{fig:T1eComp}, where the ES temperature perturbation
does not fall to zero at the ends of the simulation domain like with the IS perturbation. A more accurate model would incorporate radial variation in the diffusion
coefficient which correctly reflects linear properties of the mode. Furthermore, the Fick's law model can not account for any small but finite particle flux correctly.

To illustrate the possibility for a flux model which can capture linear mode properties, the time average of $\hat{\Gamma}_{0,\textrm{ETG}}(v_{\parallel},\mu)$ from
run $\#1$ is shown over both the linear phase and the saturated state of the nonlinear phase in Figs. \ref{fig:LinearContour} and \ref{fig:NonlinearContour}. Clearly
there is qualitative agreement between the linear and nonlinear phases, and input from linear simulations at multiple radial locations can be used to account for radial
changes in ETG mode properties. Quasilinear estimates for the flux spectra are used to better capture radial changes in ETG transport levels for each $k_y$ mode in linear
simulations:
\cite{qlTheory_Fable2010,qlTheory_Lapillonne2011}
\begin{equation}\label{qlFluxes}
   \Gamma_{k_y}^{\textrm{QL}} = A_0\frac{(\gamma/\langle k_{\perp}^2\rangle)^2}{\vert\phi_{0,k_y}(0)\vert^2}\Gamma_{k_y}^{\textrm{lin}},\;\;\;\;\;
   Q_{k_y}^{\textrm{QL}} = A_0\frac{(\gamma/\langle k_{\perp}^2\rangle)^2}{\vert\phi_{0,k_y}(0)\vert^2}Q_{k_y}^{\textrm{lin}}.
\end{equation}
$A_0$ is a constant of proportionality used to match the nonlinear fluxes and $\Gamma_{k_y}^{\textrm{lin}}$ and $Q_{k_y}^{\textrm{lin}}$ are the linear simulation
fluxes at the final time step. The electrostatic potential, $\phi_{0,k_y}(0)$, is taken at $k_x = z = 0$, and $\langle k_{\perp}^2\rangle$ is the ballooning-mode-averaged
perpendicular wavenumber squared. The flux spectrum models defined by \eqref{qlFluxes} are the same, and so can be applied to the linear kinetic flux density directly.

The quasilinear kinetic flux density is calculated at $r/a=0.50$ and shown in Fig. \ref{fig:QuasilinearContour} in comparison to the nonlinear results of Figs.
\ref{fig:LinearContour} and \ref{fig:NonlinearContour}. All parameters excluding $A_0$ are taken from linear simulations at different radial locations $r/a = 0.20,0.30,
0.40,0.50,0.60,0.70,$ and $0.80$. The parameter $A_0$ is chosen to match the nonlinear heat flux at $r/a=0.50$. While this quasilinear model has been validated for ITG
and TEM modes, it doesn't capture the ETG flux spectra as accurately; however, Figs. \ref{fig:LinearContour} - \ref{fig:QuasilinearContour} show it works well as an
initial test of feasibility. A proper quasilinear model for ETG flux spectra is beyond the scope of this work.

\begin{figure}
   \centering
   \includegraphics[width=3in]{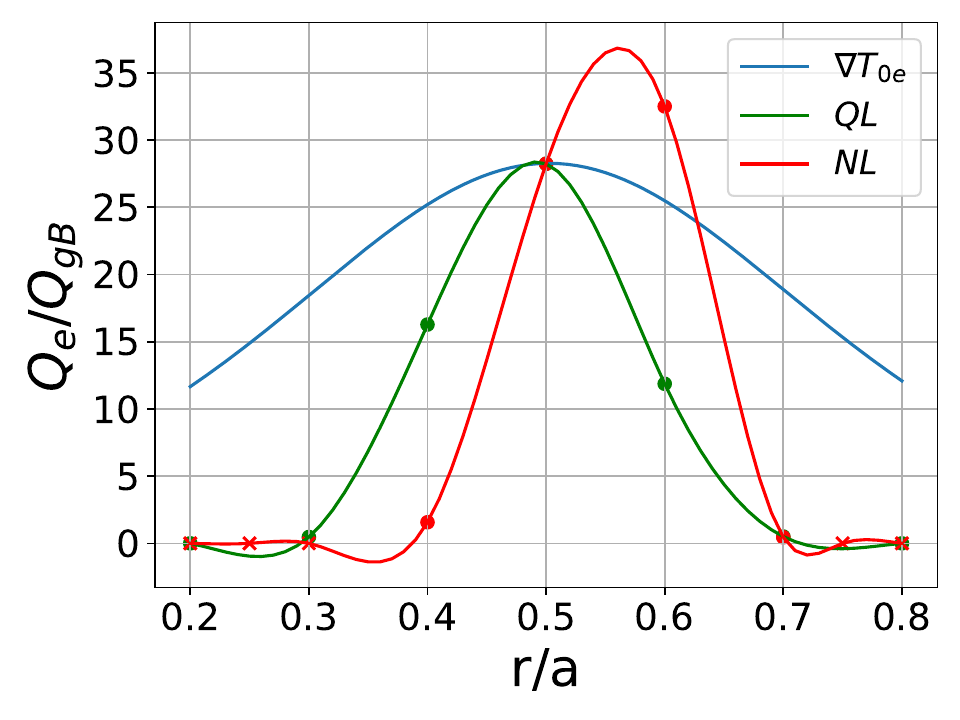}
   \caption{\label{fig:QeRadComp}Comparison of three radial ETG heat flux models with increasing fidelity. `QL' and `NL' stand for quasilinear and nonlinear respectively.
            A red x corresponds to points assumed to go to zero for nonlinear simulations.}
\end{figure}

The quasilinear model combines a single nonlinear simulation with multiple linear simulations, enabling a more efficient approach which can help expedite direct coupling
of global IS and local ES simulations. Nonlinear simulation results at $r/a = 0.40,0.50,0.60,$ and $0.70$, as described at the end of section \ref{sec:gene}, are used
to provide the most realistic radial profile of the ETG heat flux. The three different heat flux models discussed are compared directly in Fig. \ref{fig:QeRadComp}.
Spline fits are made for the quasilinear and nonlinear models using simulation results at each plotted point. All three models coincide at the peak temperature gradient
location, $r/a=0.50$, where one nonlinear simulation result must be used.

These heat flux models correspond to three different levels of fidelity. Clearly the Fick's law $\nabla T_{0e}$ model does not fall off to zero like the nonlinear results,
while the quasilinear model provides an efficient and more reasonable prediction of the radial variation for the nonlinear ETG heat flux profile. Various fits can then be
tested to include the quasilinear or nonlinear source profiles in GEM. While the focus of this work has been on capturing ES electron heat transport, the nonlinear model
would best account for any fine-scale particle transport as well.

\section{Discussion\label{sec:disc}}

A subgrid ETG model has been derived here which averages local electron-scale turbulence over intermediate scales in space perpendicular to the field and in time to
include in global ion-scale simulation. CBC simulations are carried out which show a clear scale separation in turbulent spectra in $k_y$. A kinetic form of the
electron-scale electron heat flux is taken from local GENE simulation and added into global GEM ITG simulations using a simple Fick's law diffusion model. Multiple ratios
of ITG to ETG turbulent heat flux levels are considered, and the effects of increased electron temperature relaxation are described. A more accurate quasilinear heat flux
model is constructed and compared against nonlinear ETG heat fluxes at multiple radial locations. Such a quasilinear model allows for the possibility of using a single
nonlinear ETG simulation at the peak temperature gradient only, which can help expedite coupling of simulations at both scales.

Future work will consider the effects of the ETG potential on the ion-scale distribution of electrons, as described by \eqref{smDiffEqn}. The diffusion coefficient,
$D_{\textrm{ES}}$, can be found by following the motion of tracer electrons in the ETG field of a local ES simulation, and a theoretical model developed to capture
radial variation of the diffusion. The ETG streamer potential can also be added directly to local GEM ion-scale simulations, assuming periodicity on ion scales, and
evolved in time to compare to the diffusive model. The effects of intermediate-scale zonal flows as shown in Fig. \ref{fig:PhixSpectra} might also be included as an extra
global radial shear parameter. Theoretically this would require breaking the scale separation hypothesis as $\langle\delta f_{\textrm{ES}}\rangle_{\textrm{m}} \ne 0$
and $\langle\delta\textbf{E}_{\textrm{ES}} \rangle_{\textrm{m}} \ne 0$ when the spectra become multiscale in $k_x$. The formation of intermediate-scale zonal flows
in multiscale simulation is dependent on the level of ITG turbulence,\cite{MultiscaleIBS_Holland2017} and so it would be prudent to first focus on including effects
of ion-scale turbulence in electron-scale simulations per Ref. \onlinecite{MultiscaleModel_Hardman2019}.

Many further topics exist for future research directions. These include adding the compressible effects and effects of ETG Reynolds stress which have been ignored here,
considering any spectral transfer between scales, and proper coupling of ITG and ETG simulations to capture effects of ion-scale turbulence in electron-scale simulations.
Furthermore, it is important to understand when a scale separation hypothesis is valid, as electron transport spectra can broaden to become multiscale in the pedestal.
\cite{MutliscalePedestal_Belli2023}

\begin{acknowledgments}
The authors would like to acknowledge Gabriele Merlo for helpful discussion about the GENE code and normalizations. This research was supported by the Frontiers in Leadership
Gyrokinetic Simulation project, Scientific Discovery through Advanced Computing program, U.S. Department of Energy Contract \#DE-SC0024425. This research utilized resources of
the National Energy Research Scientific Computing Center (NERSC), a Department of Energy Office of Science User Facility, through NERSC award FES-ERCAP0026751.
\end{acknowledgments}

\appendix

\section{Normalizations\label{sec:norm}}

The normalizations involved in converting GENE output to GEM input are discussed here. Including \eqref{SrcETG2} in GEM requires converting the amplitude as well as
the velocity-space grid to be consistent at different radii with global variation in temperature. The conversion from normalized units to SI units for parallel
velocity and magnetic moment are defined in each code as\cite{Gem2_Chen2007,MerzThesis}
\begin{equation}\label{vparNorms}
   v_{\parallel,e}^{\textrm{SI}} = v_{\parallel,e}^{\textrm{GN}}\sqrt{\frac{2T_{0e}(r)}{m_e}},\;\;\;
   v_{\parallel,e}^{\textrm{SI}} = v_{\parallel,e}^{\textrm{GM}}\sqrt{\frac{T_{0e}(r_0)}{m_p}},
\end{equation}
\begin{equation}\label{muNorms}
   \mu_e^{\textrm{SI}} = \mu_e^{\textrm{GN}}\frac{T_{0e}(r)}{B_0},\;\;\;
   \mu_e^{\textrm{SI}} = \mu_e^{\textrm{GM}}\frac{T_{0e}(r_0)}{B_0}.
\end{equation}
`GN' and `GM' stand for GENE and GEM respectively. Eqs. \ref{vparNorms} and \ref{muNorms} are used to convert the normalized velocity-space grid in GENE to SI values
at multiple radial locations, and then to GEM normalized units. The velocity-space grid is recalculated for each particle depending on its radial location and particles
are interpolated onto the new velocity-space grid to calculate $\hat{\Gamma}_{\textrm{ETG}}$. Approximately $95\%$ of GEM particles fall into the velocity-space grid at all
radial locations, with the remaining particles outside the $v_{\parallel,e}$ domain $\pm3 v_{Te}$ set in GENE, where $v_{Te} = \sqrt{2T_{0e}(r)/m_e}$. 

Furthermore, the magnitude of $\hat{\Gamma}_{0,\textrm{ETG}}$ needs to be converted to GEM normalized units. The quantities $v_{Ex}$ and $\delta f_e$ are normalized as
follows
\begin{equation}
   v_{Ex}^{\textrm{SI}} = \rho^*\sqrt{\frac{T_{0e}(r_0)}{m_D}}v_{Ex}^{\textrm{GN}},\;\;\;
   v_{Ex}^{\textrm{SI}} = \sqrt{\frac{T_{0e}(r_0)}{m_p}}v_{Ex}^{\textrm{GM}},
\end{equation}
\begin{equation}
\begin{aligned}
   \delta f_e^{\textrm{SI}} &= \delta f_e^{\textrm{GN}}\rho^*n_{0e}(r)/v_{Te}^3(r),\;\;\; \\
   \delta f_e^{\textrm{SI}} &= \delta f_e^{\textrm{GM}}n_{0e}(r_0)/(T_{0e}(r_0)/m_p)^{3/2},
\end{aligned}
\end{equation}
where GENE includes a factor of $\rho^* = \rho_D/R$ scaling for the perturbations $\delta\phi$ and $\delta f$. Radial variation in $v_{Ex}$ is ignored to use only the peak
turbulence level. While GENE uses the radial basis vector $\textbf{e}_x = \nabla r$ by default, GEM uses the unit vector $\hat{\textbf{e}}_x = \nabla_r/\vert\nabla_r\vert$ for the
radial dot product. For the circular geometry used this makes no difference, however this can be changed in GENE using the `norm\_flux\_projection' flag if necessary.

\begin{figure}
   \centering
   \subfigure[]{
   \label{fig:SourceComparison}
   \includegraphics[width=3in]{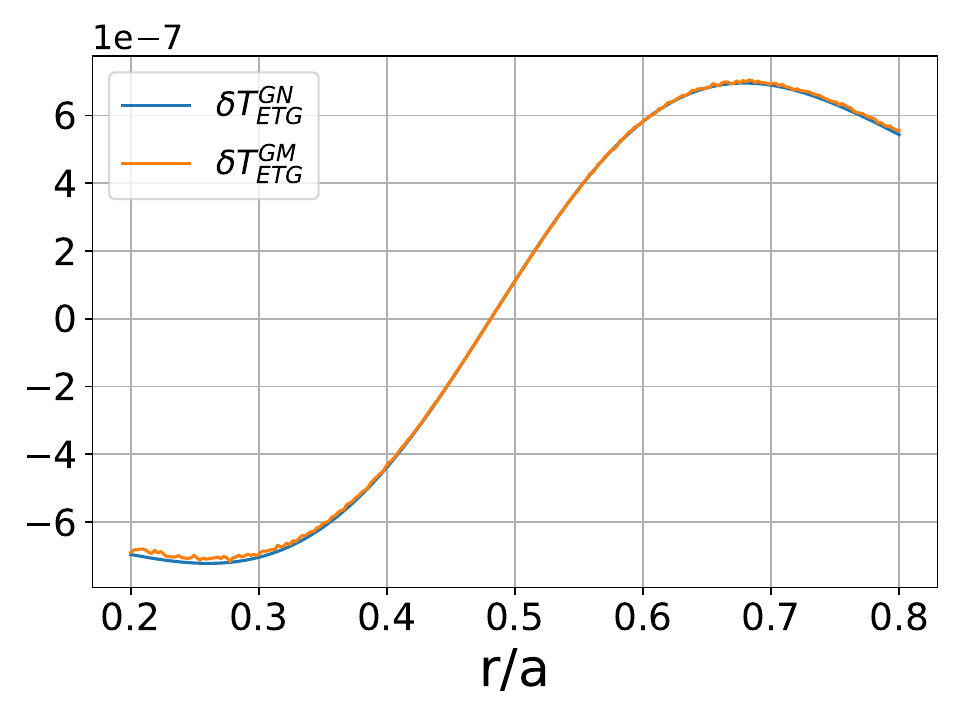}}
   \quad
   \subfigure[]{
   \label{fig:FluxTerms}
   \includegraphics[width=3in]{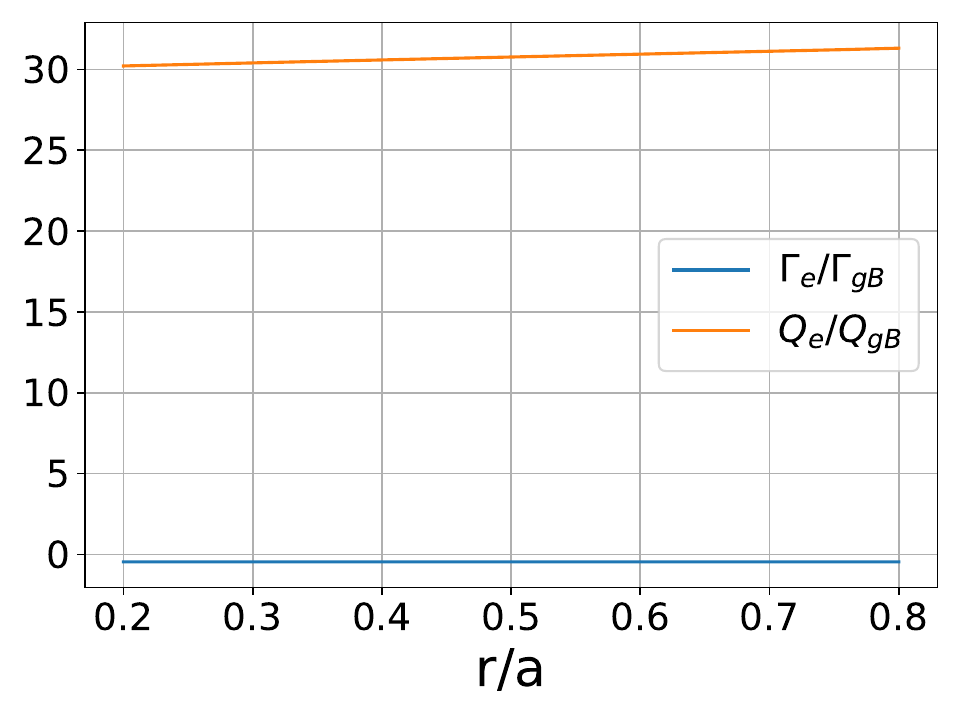}}
   \caption{\label{fig:TestPlots} (a) Comparison of source terms when integrating GENE data directly in velocity-space and when integrating by interpolating over particles
            in GEM. (b) Electron flux profiles when $\hat{\Gamma}_{0,\textrm{ETG}}$ is integrated assuming only radial changes in geometry ($B(r,z),J(r,z)$).}
\end{figure}

Finally, $L_{\textrm{ref}}^{\textrm{GN}} = R$, while $L_{\textrm{ref}}^{\textrm{GM}} = \rho_p$, the proton gyroradius, so that the normalized second derivative,
\eqref{grad2T0}, must be multiplied by a factor $(\rho_p/R)^2$, giving a total factor
\begin{equation}\label{srcConv}
   \nabla\cdot\hat{\Gamma}_{\textrm{ETG}}^{\;\textrm{GM}} = \frac{n_{0e}(r)}{n_{0e}(r_0)}\left(\frac{2T_{0e}(r)}{T_{0e}(r_0)}\frac{m_p}{m_e}\right)^{-3/2}
   (\frac{\rho_p}{R})^2(\frac{\rho_D}{R})^2(\nabla\cdot\hat{\Gamma}_{\textrm{ETG}}^{\;\textrm{GN}}).
\end{equation}
While this correctly accounts for differences in normalizations between the codes it does not yet give a radially constant $\hat{D}^*_{0,\textrm{ETG}}$ consistent with \eqref{SrcETG2}.
This is because \eqref{srcConv} depends radially on $n_{0e}(r)$. The other factor of $T_{0e}^{-3/2}(r)$ is cancelled by \eqref{vparNorms} and \eqref{muNorms} when integrating over
velocity-space. Furthermore, when calculating the heat flux an extra $\frac{1}{2}mv^2$ factor will add another radial dependence on $T_{0e}(r)$. To see this, consider the GENE formula
for the normalized heat flux\cite{MerzThesis}
\begin{equation}\label{geneQe}
\begin{aligned}
   \frac{\langle Q_e\rangle_{\psi}}{Q_{gB}} = -\frac{\hat{n}_{0e}(r)\hat{T}_{0e}(r)}{\int_{-\pi}^{\pi}\hat{J}(r,z)dz}&\int_{-\pi}^{\pi}\sum_{\textbf{k}}\hat{J}(r,z)i\hat{k}_y
   \delta\hat{\bar{\phi}}(\textbf{k}) \\
   &\times\left(\pi \hat{B}_0(r,z)\int d\hat{v}_{\parallel}d\hat{\mu} \hat{v}^2\delta \hat{f}_e(\textbf{k})\right)^*,
\end{aligned}
\end{equation}
which contains radial dependence on $\hat{p}_{0e}(r)=\hat{n}_{0e}(r)\hat{T}_{0e}(r)$. A caret ($\textrm{\textasciicircum}$) in \eqref{geneQe} indicates normalized variables.

The extra pressure factor is divided from \eqref{srcConv} to retrieve an approximately radially-constant $\hat{D}^*_{0,\textrm{ETG}}$. The source term is integrated on
the original GENE velocity-space grid and converted to GEM normalization to compare to direct integration in GEM which sums over particle weights. The results of both
integration methods are compared in Fig. \ref{fig:SourceComparison} using the kinetic flux density of Fig. \ref{fig:MesoscaleContour} and agree well. Note, some radial
variation remains in $\hat{D}^*_{0,\textrm{ETG}}$ due to radial changes in $J(r,z)$ and $B_0(r,z)$ which are used when integrating \eqref{geneQe}. The effect of this is
small and simply ignored. This remaining radial variation can be seen in Fig. \ref{fig:FluxTerms}, where the fluxes are calculated by integrating $\hat{\Gamma}_{0,
\textrm{ETG}}$ in Fig. \ref{fig:MesoscaleContour} using CBC equilibrium profiles.

\nocite{*}
\bibliography{SubgridETG}

\end{document}